\documentclass[twocolumn]{aastex631}

\usepackage{bm}
\usepackage{amsmath}
\usepackage{dblfloatfix}
\usepackage{breqn}

\newcommand\Kepler{\textit{Kepler}}

\newcommand\ms{$\textrm{m~s}^{-1}$}

\newcommand\gcmcubed{$\textrm{g~cm}^{-3}$}

\newcommand\earthmass{$M_{\oplus}$}
\newcommand\earthradius{$R_{\oplus}$}
\newcommand\earthinsol{$S_{\oplus}$}

\newcommand\solmass{$M_{\odot}$}

\hypersetup{citecolor=teal}

\begin{document}

\title{Beyond 2-D Mass-Radius Relationships: A Nonparametric and Probabilistic Framework for Characterizing Planetary Samples in Higher Dimensions}


\author[0000-0001-8401-4300]{Shubham Kanodia}
\affiliation{Earth and Planets Laboratory, Carnegie Institution for Science, 5241 Broad Branch Road, NW, Washington, DC 20015, USA}

\author[0000-0002-5223-7945]{Matthias Y. He}
\affiliation{Department of Physics \& Astronomy, 225 Nieuwland Science Hall, The University of Notre Dame, Notre Dame, IN 46556, USA}

\author[0000-0001-6545-639X]{Eric B.\ Ford}
\affiliation{Department of Astronomy \& Astrophysics, 525 Davey Laboratory, The Pennsylvania State University, University Park, PA, 16802, USA}
\affiliation{Center for Exoplanets \& Habitable Worlds, 525 Davey Laboratory, The Pennsylvania State University, University Park, PA, 16802, USA}
\affiliation{Institute for Computational and Data Sciences, The Pennsylvania State University, University Park, PA, 16802, USA}
\affiliation{Center for Astrostatistics, 525 Davey Laboratory, The Pennsylvania State University, University Park, PA, 16802, USA}

\author[0000-0001-8351-408X]{Sujit K. Ghosh}
\affiliation{Department of Statistics, 2311 K. Stinson Drive, North Carolina State University, Raleigh, NC 27695-8203, USA}

\author[0000-0003-2862-6278]{Angie Wolfgang}
\affiliation{Eureka Scientific, Inc., 2452 Delmer Street Suite 100, Oakland, CA, 94602-3017, USA}

\correspondingauthor{Shubham Kanodia}
\email{skanodia@carnegiescience.edu}

\begin{abstract}
Fundamental to our understanding of planetary bulk compositions is the relationship between their masses and radii, two properties that are often not simultaneously known for most exoplanets. However, while many previous studies have modeled the two-dimensional relationship between planetary mass and radii, this approach largely ignores the dependencies on other properties that may have influenced the formation and evolution of the planets.
In this work, we extend the existing nonparametric and probabilistic framework of \texttt{MRExo} to jointly model distributions beyond two dimensions.
Our updated framework can now simultaneously model up to four observables, while also incorporating asymmetric measurement uncertainties and upper limits in the data.
We showcase the potential of this multi-dimensional approach to three science cases: (i) a 4-dimensional joint fit to planetary mass, radius, insolation, and stellar mass, hinting of changes in planetary bulk density across insolation and stellar mass; (ii) a 3-dimensional fit to the California Kepler Survey sample showing how the planet radius valley evolves across different stellar masses; and (iii) a 2-dimensional fit to a sample of Class-II protoplanetary disks in Lupus while incorporating the upper-limits in dust mass measurements.
In addition, we employ bootstrap and Monte-Carlo sampling to quantify the impact of the finite sample size as well as measurement uncertainties on the predicted quantities. We update our existing open-source user-friendly \texttt{MRExo} \texttt{Python} package with these changes, which allows users to apply this highly flexible framework to a variety of datasets beyond what we have shown here.
\end{abstract}



\section{Introduction} \label{sec:intro}
In the $\sim30$ years since the discovery of the first extra-solar planets \citep{wolszczan_planetary_1992, mayor_jupiter-mass_1995}, astronomers have discovered over 5000 exoplanets \citep[NASA Exoplanet Archive;][]{akeson_nasa_2013}. The growth in sample size lends itself to the use of increasingly sophisticated statistical tools and increasing the dimensionality of the models used for interpreting the exoplanet sample (and population). For example, based on the first handful of known exoplanets from radial velocities (RVs), \cite{gonzalez_stellar_1997} noted a preference for giant planets to be found around metal-rich host stars. This trend has held up with more sophisticated analysis of larger samples of short-period giant exoplanets from both RV and transiting surveys \citep{santos_metal-rich_2001, fischer_planet-metallicity_2005, ghezzi_stellar_2010, sousa_spectroscopic_2011, buchhave_three_2014, wang_revealing_2015, petigura_california-keplersurvey_2018, narang_properties_2018}. 

Another observed feature in the exoplanet population is the ``radius gap'' for small exoplanets ($R_p <$ 4 \earthradius) that was first predicted by \cite{owen_kepler_2013} and \cite{lopez_role_2013} and identified observationally by \cite{fulton_california-kepler_2017} using a sample of transiting planets from the \textit{Kepler} mission \citep{borucki_kepler_2010} combined with precise stellar parameters from the California-\textit{Kepler} Survey \citep[CKS; ][]{petigura_california-kepler_2017}.  Initially, the radius gap referred to a deficit in planets with radii $\sim 1.7$ \earthradius\, in a histogram (1-D space).  Now, it refers to a valley in 2-D planet radius-orbital period space \citep{fulton_california-kepler_2017, van_eylen_asteroseismic_2018, berger_revised_2018, martinez_spectroscopic_2019, 2019AJ....158..109H}. In a quest to disambiguate between the various physical mechanisms that can produce this deficit of planets, astronomers have considered how the radius gap varies with additional planetary and stellar parameters, e.g., using slices of the 1-D radius histogram or 2-D radius-period plane for different stellar properties \citep[e.g.,][]{fulton_california-kepler_2018, berger_gaia-kepler_2020, van_eylen_masses_2021, otegi_revisited_2020}, stellar metallicities \citep{owen_metallicity-dependent_2018, petigura_california-kepler_2022, otegi_revisited_2020}, and ages \citep{berger_gaia-kepler_2020, petigura_california-kepler_2022}.

Similar 2-D models have been used in a wide variety of exoplanet studies, such as planet mass-metallicity relations \citep[][and references therein]{welbanks_mass-metallicity_2019} and the haziness of planet atmospheres \citep[e.g.,][]{yu_haze_2021, edwards_exploring_2022}. Likewise, ALMA measurements of Type II protoplanetary disks have helped estimate the mass in dust (with continuum measurements at $\sim 870~\mu$m) and gas (typically using CO and its isotopologues).  These studies found the relationships depend on the host stellar mass \citep[e.g.,][]{andrews_mass_2013, ansdell_alma_2016, pascucci_steeper_2016} and other stellar properties. 

Of course, planet structure models rely on more than two parameters, typically including 4-5 dimensions -- planet radius, mass, equilibrium temperature (or insolation flux), and age -- \citep{fortney_planetary_2007, baraffe_structure_2008, muller_synthetic_2021}.
A similar increase in the dimensionality and the complexity of modeling tools has also taken place for characterizing planetary mass-radius (MR) relations.  Initially, studies assumed deterministic 2-D power laws \citep{seager_mass-radius_2007, wu_density_2013, weiss_mass-radius_2014, thorngren_empirical_2019}.  More recently, studies have used Hierarchical Bayesian Modeling (HBM) to develop probabilistic models based on a 2-D power-law  \citep{wolfgang_probabilistic_2016} or piecewise power-law over the mass-radius plane \citep{bashi_two_2017, chen_probabilistic_2017, otegi_revisited_2020}. 
These parametric models assume a relatively simple mathematical model over some region of the MR plane. However, it appears that a more flexible model is required to capture the MR relation over a broader range of planetary radii and masses. 
Further, most of these models cannot reproduce all of the observed features in the 2-D period-radius ($P$-$R_p$) plane, such as the radius-valley or the Neptune desert \citep{mazeh_dearth_2016}. Additionally, it is not clear what the functional form of these relations should be, particularly  when additional dimensions beyond just mass and radius are considered. 

In parallel, a variety of nonparametric methods have been employed (e.g., beta density functions, \citealt{ning_predicting_2018, kanodia_mass-radius_2019}; the Maximum Entropy approach, \citealt{ma_maximum_2019}; random forests, \citealt{ulmer-moll_beyond_2019}; neural networks, \citealt{tasker_estimating_2020}) to characterize the exoplanet MR relation. MR relations can be useful to infer the composition of planets based on their bulk density \citep{lopez_understanding_2014, rogers_most_2015, zeng_growth_2019} and for predicting other planetary properties \citep{chen_probabilistic_2017, kanodia_mass-radius_2019}. Some studies have expanded the MR relationship to three dimensions (MR+) either using a product of power laws \citep{weiss_mass_2013} or Bayesian models \citep{neil_host_2018, neil_joint_2020, ma_orbital_2021}.

In this work we expand on previous work by \cite{ning_predicting_2018} and \cite{kanodia_mass-radius_2019} offering a nonparametric method for inferring the probability density describing a 2-D sample using beta density functions. Here we allow for the simultaneous modeling of up to four dimensions\footnote{While the current algorithm can fit four dimensions, it can be trivially expanded to higher dimensions if required.} and provide an implementation in the updated \texttt{MRExo}\footnote{\url{https://github.com/shbhuk/mrexo}} Python package \citep{kanodia_mass-radius_2019}. While primarily developed as an expansion to the MR relation, it can be used as a general purpose modeling tool between any (up to four) measured quantities.  Additionally, it has been generalized to work with symmetric or asymmetric measurement uncertainties and can incorporate observations resulting in upper limits. Some examples of 3-D spaces that can be modeled using such a framework are: studying Type II disk dust mass (including upper limits) as a function of stellar mass and age, estimating log(\textit{C/O}) as a function of planet radius and insolation flux. Similarly in 4-D, one can jointly model the M-R-insolation space as a function of stellar mass, or conversely M-R and orbital separation as a function of stellar metallicity. \texttt{MRExo} can also be used to infer the dependence of  water scale height in transmission spectroscopy (or conversely haze amplitude) as a function of equilibrium temperature, surface gravity and stellar insolation (bolometric or high-energy).

In Section \ref{sec:model} we describe the generalized nonparametric model.  In Section \ref{sec:applications} we present a few scientific applications to demonstrate the utility and advantage of multi-dimensional nonparametric approach.  Finally, we conclude in Section \ref{sec:conclusion}. A detailed appendix discusses the updates to the model from previous work, as well as some salient features of the model.

\section{Nonparametric Model}\label{sec:model}

We expand the framework from \cite{ning_predicting_2018} and \cite{kanodia_mass-radius_2019} to use Bernstein polynomials\footnote{When normalized, each Bernstein polynomial has the same functional form as beta density functions.  See the Appendix from \cite{ning_predicting_2018} for more details on the choice of basis function.} for multivariate density estimation, i.e. to model an $n$-dimensional joint distribution of the variables $f(x_1, x_2,..,x_n)$, where $x_t$ represents the variable within different dimensions such as mass, radius, insolation, etc. 
This is essentially the probability of having a particular set of variables $x_1, x_2,..,x_n$, given weights (or coefficients) $\bm{w}$, and polynomial degrees $d^{(1)},...,d^{(n)}$. 

\begin{figure}[] 
\centering
\includegraphics[width=0.5\textwidth]{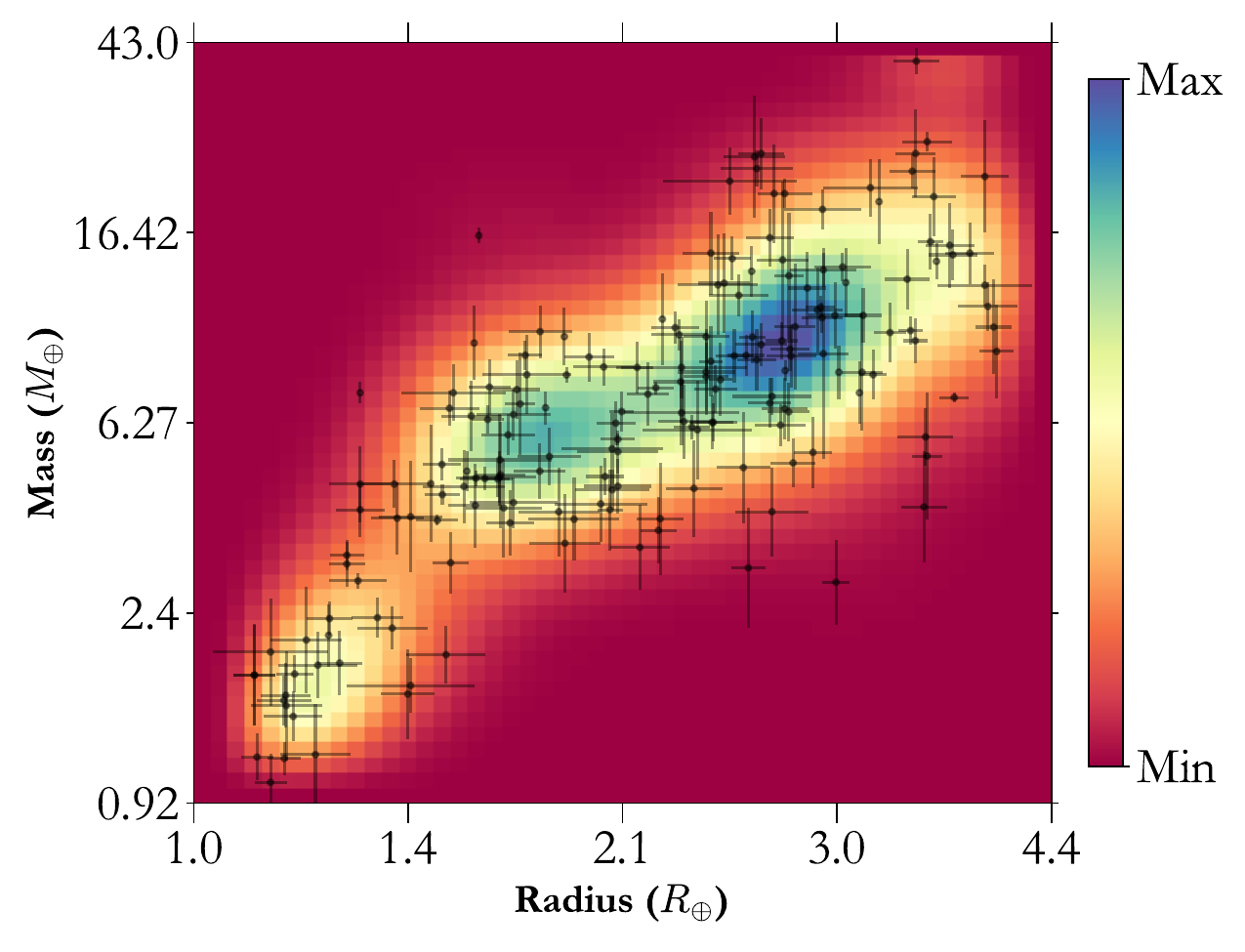}
\caption{2-D sample set with $\sim$ 180 planets. Joint distribution $f(m, r)$ showing the masses and radii for the input dataset, where the colours in the background represent the PDF, where blue is a higher probability and red is lower.} \label{fig:joint}
\end{figure}

For example, if we assume a 3-D joint distribution, $f(x_1, x_2, x_3| \bm{w}, d^{(1)}, d^{(2)}, d^{(3)}))$, for the probability density of a planet to have three properties $x_1$, $x_2$, and $x_3$, then the value of $f$ at each point in this continuous space should be interpreted as the probability for a planet to exist with given $x_1$, $x_2$, $x_3$. 
The density is uniquely specified by a set of non-negative weights $\bm{w}$ and degrees $d^{(1)}, d^{(2)}, d^{(3)}$.
Next, to use this model for predictive purposes, the joint distribution can be conditioned to obtain the probability density function (PDF). Following the laws of conditional probability (see Equation 10 from \citealt{ning_predicting_2018}), 
\begin{align}
        &f(x_1|x_2,x_3, \bm{w}, d^{(1)}, d^{(2)}, d^{(3)}) \notag \\ 
        &= \frac{f(x_1, x_2, x_3 | \bm{w}, d^{(1)}, d^{(2)}, d^{(3)})}{\int f(x_1, x_2, x_3 | \bm{w}, d^{(1)}, d^{(2)}, d^{(3)}) ~ \textrm{d}x_1}  \\ 
        &= \frac{f(x_1, x_2, x_3 | \bm{w}, d^{(1)}, d^{(2)}, d^{(3)})}{f(x_2, x_3 | \bm{w}, d^{(1)}, d^{(2)}, d^{(3)})},
\end{align}

Then, the expected value  for $x_1$ can be computed from the PDF,
\begin{align}
        & E(x_1) = \frac{\int x_1 f(x_1 | ...)~ \textrm{d}x_1}{\int f(x_1 | ...)~ \textrm{d}x_1} \Rightarrow \int x_1 f(x_1 | ...)~ \textrm{d}x_1. \label{eq:Expectation}
\end{align}

One advantage of the Bernstein polynomial formulation is that conditional probabilities and expectation values can be computed efficiently.  
The detailed mathematical formalism for the general $n$-dimensional case is included in the Appendix \ref{sec:generalizing}, including the joint distribution (Appendix \ref{sec:joint}) and the likelihood for the model (Appendix \ref{sec:calc_likelihood}). The likelihood is maximized to estimate the two unknown sets of parameters in the model, the matrix of weights $\bm{w}$, and the choice of degrees for each dimension (Appendix \ref{sec:optimize_likelihood}). Similar to \cite{kanodia_mass-radius_2019}, we set the edge weights in each dimension (e.g., first and last row and column in 2-d) to zero to reduce edge effects. Furthermore, we allow for the possibility of asymmetric measurement uncertainties which allows us to include upper (or lower) limits in our framework. We incorporate this into \texttt{MRExo} by modifying the framework as explained in Appendix \ref{sec:asymmetric}. Lastly, we include the possibility to estimate the optimum number of degrees for a given dataset by either using the Akaike Information Criterion \citep[AIC;][]{akaike_information_1973} or $k$-fold cross-validation \citep[CV;][]{james_introduction_2013} and discuss this further in Appendix \ref{sec:degree_selection}.

\begin{figure}[!t] 
\centering
\includegraphics[width=0.5\textwidth]{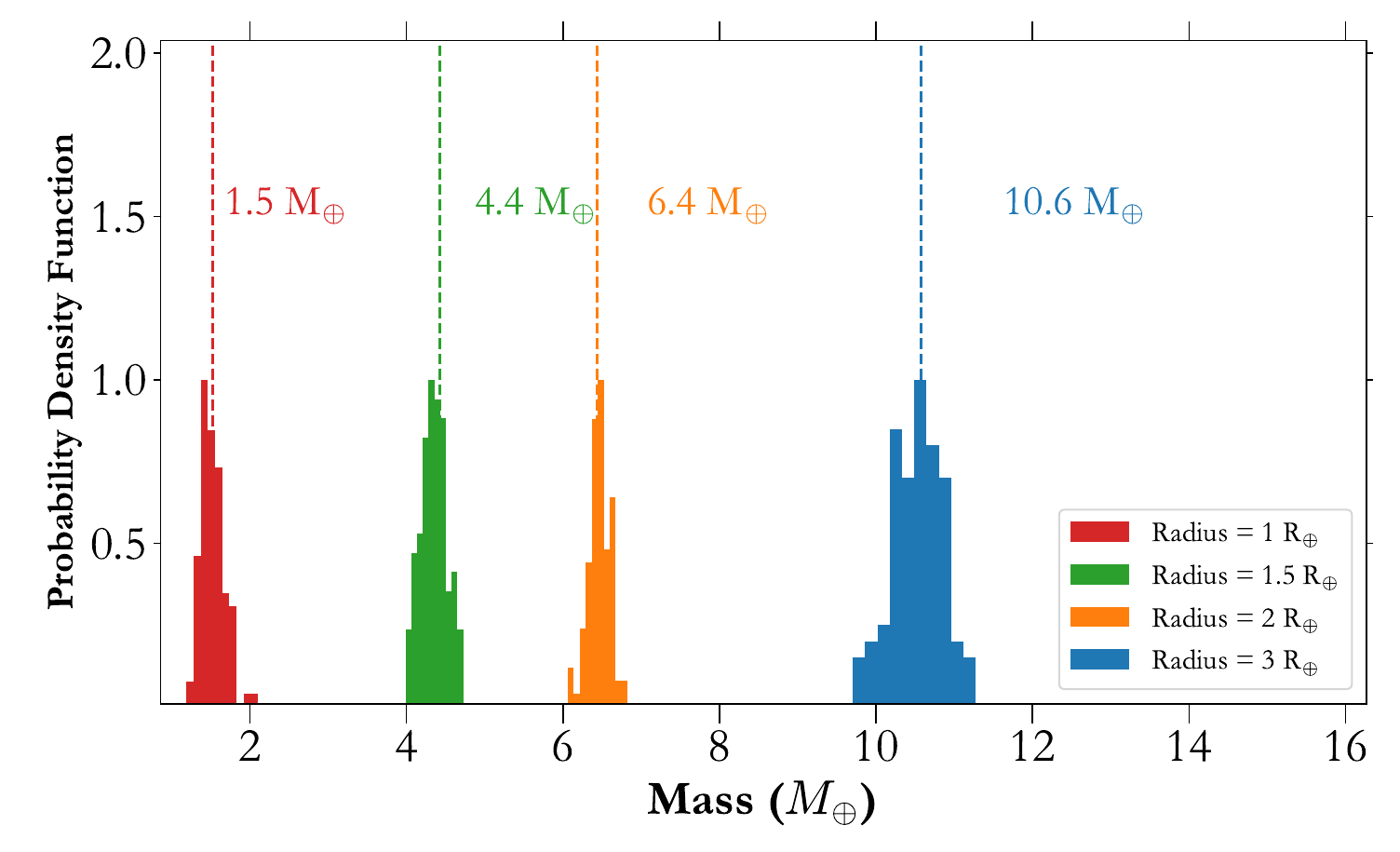}
\caption{Conditional mass distribution $f(m|r)$ for the joint distribution shown in \autoref{fig:joint} on three different radii to obtain the conditional distribution for masses. The dashed lines and the masses are the expected value (\autoref{eq:Expectation}) for each radius, whereas the histogram shows the predictions from each Monte-Carlo for illustrative purposes. } \label{fig:conditional}
\end{figure}

\begin{figure}[ht]
\fig{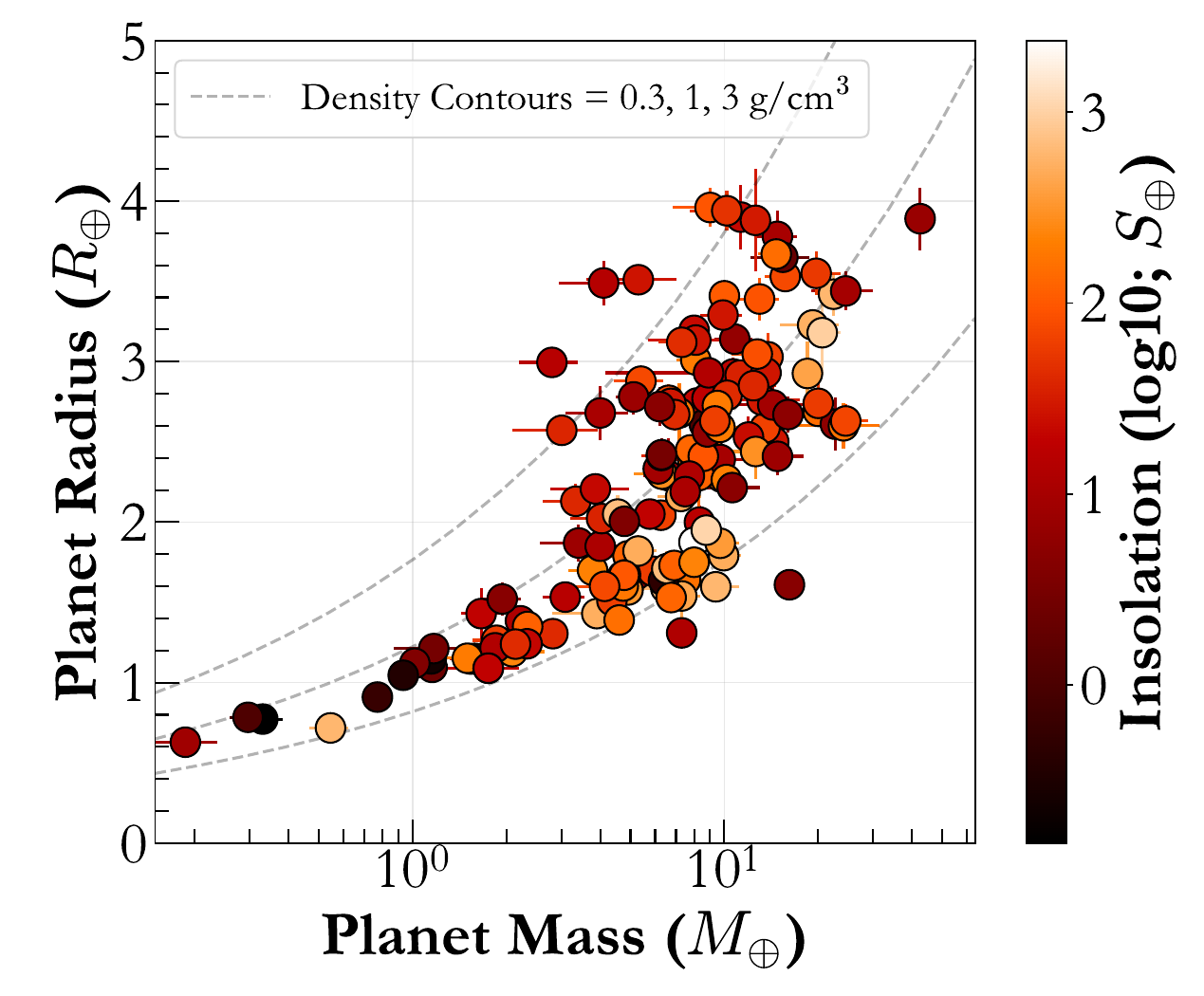}{0.45\textwidth}{}
\fig{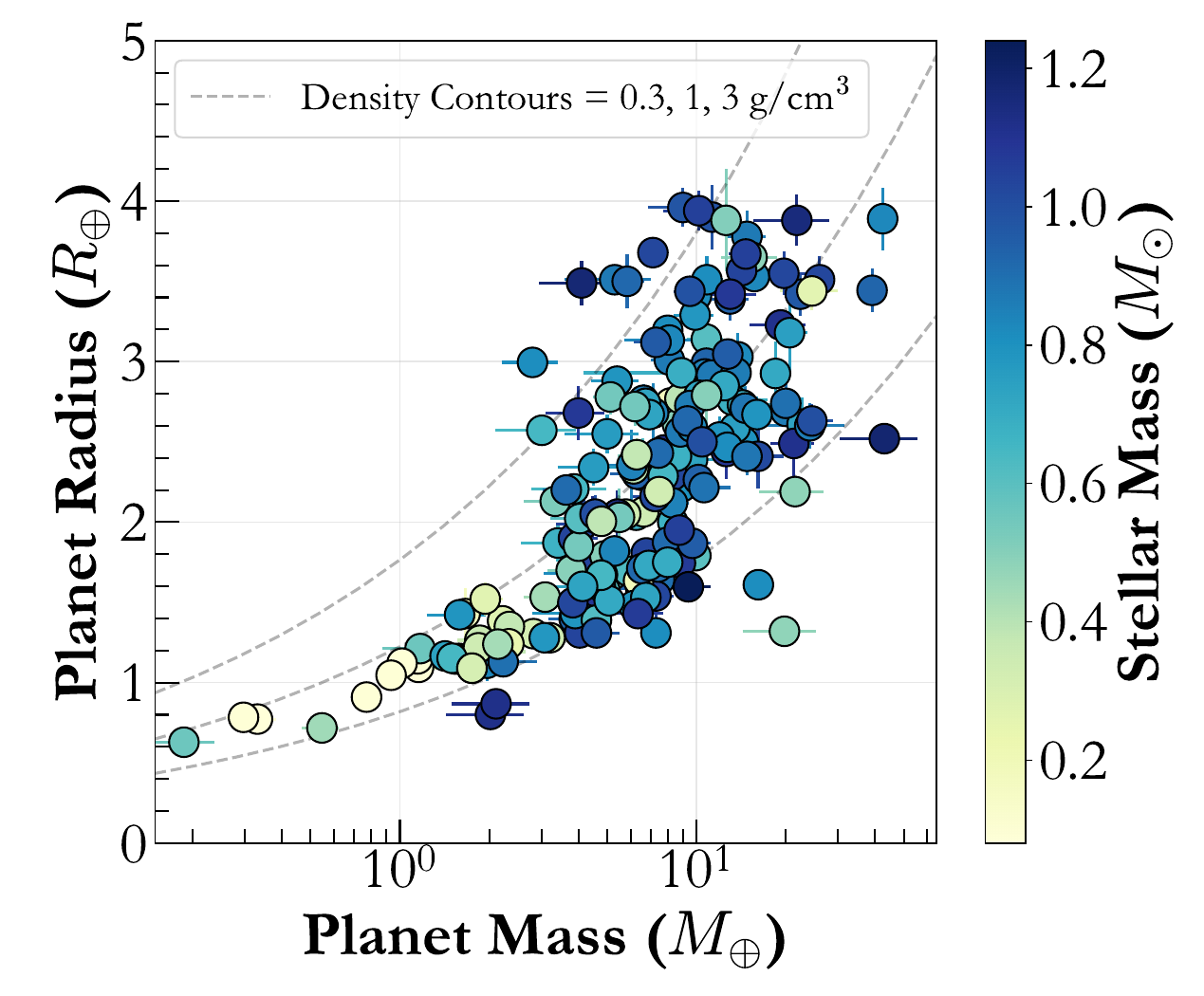}{0.45\textwidth}{}
\caption{\small 2-dimensional representations of the input sample in planetary mass-radius plane coloured by stellar insolation and stellar mass respectively. It is evident that insolation and/or stellar mass dependent bulk-density trends are harder to tease out with conventional 2-dimensional analysis.}\label{fig:MR2d}
\end{figure}

For illustrative purposes, we show an example of a MR joint distribution --- $f(m,r)$ --- in 2-D for a sample of 182 planets from the NASA Exoplanet Archive \citep{akeson_nasa_2013, PSCompPars} queried on 2023 March 6 for planets with masses and radii $> 3\sigma$ precision, planetary radii $<$ 4 \earthradius~ and stellar masses $<$ 1.5 \solmass.
The data and fitted joint distribution in planet mass-radius are shown in Figure \ref{fig:joint}.
We used \texttt{MRExo} with this sample of $\sim$ 180 planets and the cross-validation method to select 40 degrees for this model by maximizing the log-likelihood as discussed in \cite{ning_predicting_2018}.
The mean predictions for the distribution of planet masses conditioned on several values of planet radii are shown in \autoref{fig:conditional}. In these figures (and throughout this paper), these distributions quantify the \textit{observed} samples instead of the intrinsic populations due to detection biases, for which a detailed treatment is outside the scope of this work.
Future work can combine the Bernstein polynomial formalism for nonparametric density estimation with other techniques to  account for detection bias for characterizing an underlying population.
Planet formation and evolution models predict that planet interiors and bulk-structure is likely to depend on the size of the host-star, the nature (and quantity) of primordial material in protoplanetary disks, stellar metallicity,  stellar luminosity, etc. \citep{ida_toward_2004_1, ida_toward_2005, fortney_planetary_2007, burn_new_2021}, which motivate a higher-dimensional extension to the mass-radius relationship. This is also seen empirically in \autoref{fig:MR2d}, by colour-coding the planet mass-radius plane by stellar mass and insolation flux.

\begin{figure*}[ht] 
\centering
\includegraphics[width=0.8\textwidth]{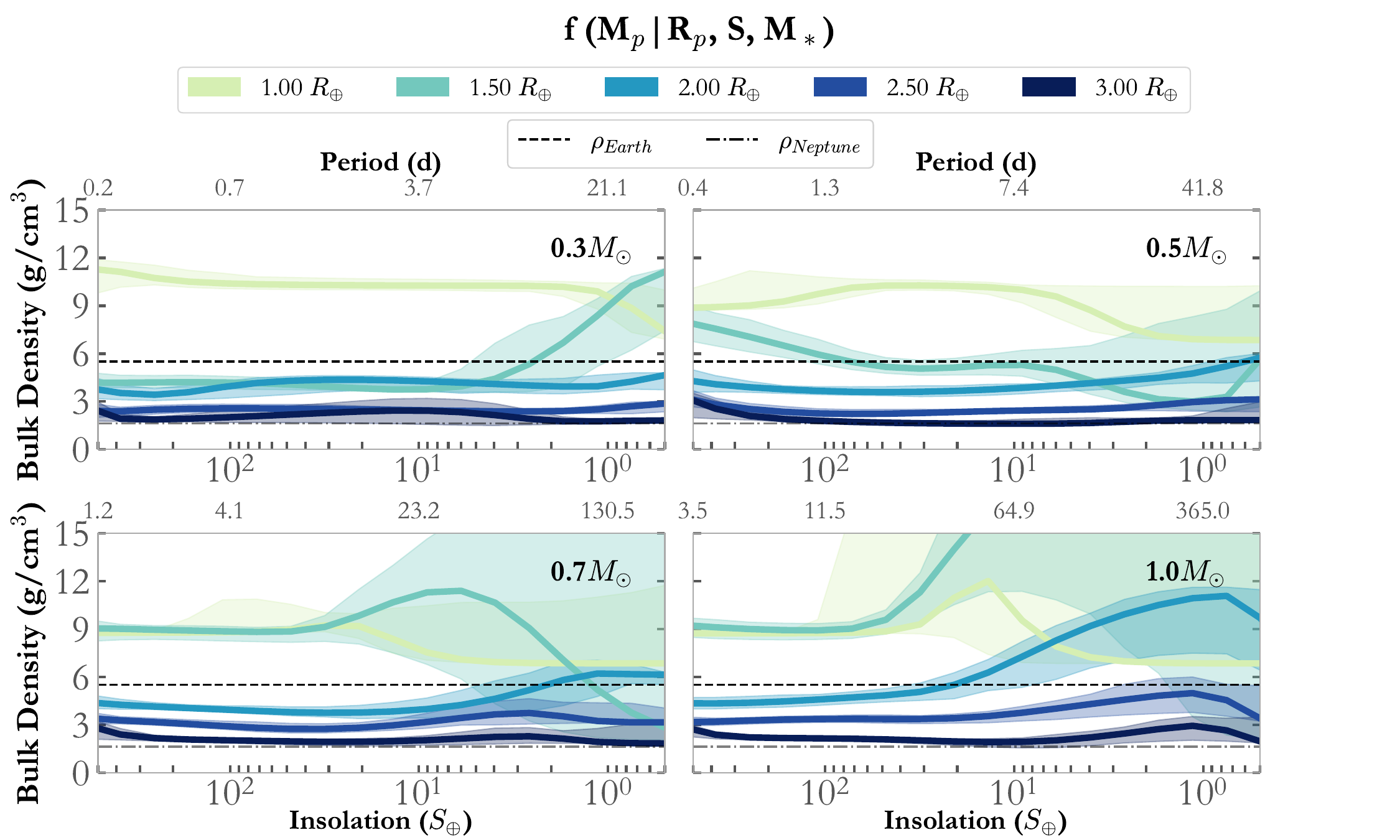}
\caption{We perform a 4-dimensional fit to the sample of confirmed small planets ($R_p < 4$\earthradius), and then condition it to show the variation in planetary bulk-density with insolation and stellar mass. For each panel, the bottom $x$-axis denotes the insolation flux received by the planet (same for all panels) while the top $x$-axis denotes the orbital period (which varies across the panels due to the different stellar masses). To convert insolation to orbital period, we adopt stellar luminosities from Table 6 in \cite{cifuentes_carmenes_2020}. The solid lines depict the expectation value (mean) of the prediction, whereas the shaded region shows the 16th -- 84th percentile region from 100 bootstrapped samples. For reference, we also plot the values for the mean bulk densities of Earth and Neptune as the dashed and dashed-dotted lines, respectively.} \label{fig:bulkdensity_stmass4d}
\end{figure*}

\begin{figure*}[ht] 
\centering
\includegraphics[width=0.8\textwidth]{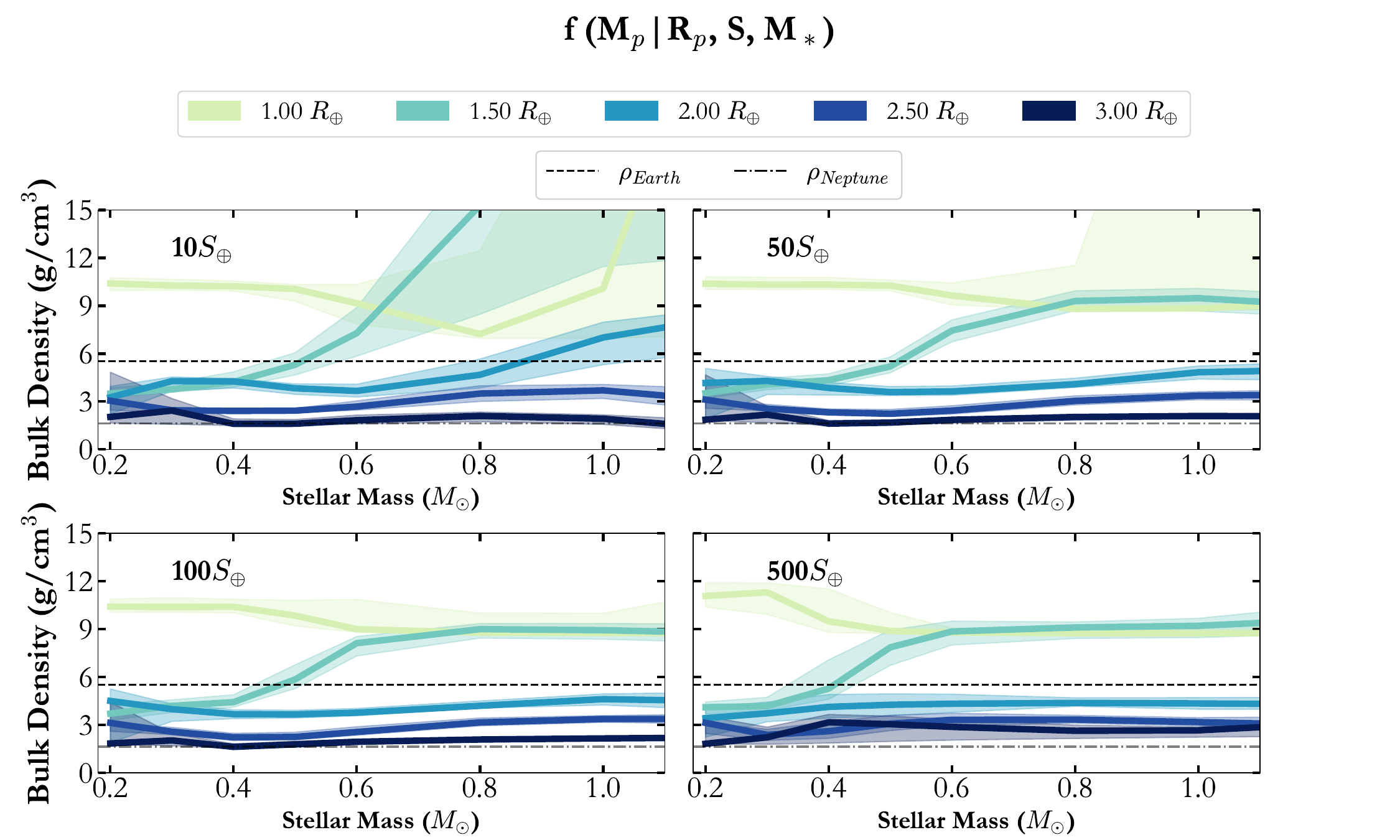}
\caption{Similar to the previous figure, we condition the 4-dimensional joint distribution to obtain the expectation value for planetary mass as a function of different planetary radii, insolation and plotted across stellar mass. The  shaded region depicts the 16th -- 84th percentile region from 100 bootstrapped samples.} \label{fig:bulkdensity_insol4d}
\end{figure*}

\section{Science applications} \label{sec:applications}
In this section we give a few examples of the updated \texttt{MRExo} framework, allowing for higher dimensional datasets and also asymmetric errorbars to include upper (or lower) limits in the models.

\subsection{Mass-Radius+ with \texttt{MRExo}}\label{sec:MR+}

We apply a 4-D model to the dataset described in Section \ref{sec:model} with $\sim$ 180 planets, across planetary masses, radii, insolation fluxes, and stellar masses. By extending the cross-validation framework described in \cite{ning_predicting_2018} to higher dimensions here, we select the optimum degrees after maximizing the log-likelihood to be 40 in each dimension. Here we note that while the framework allows for unequal degrees in each dimension, for speed and simplicity, we assume an equal number of degrees. This results in $(40-2)^4 \simeq 2.1 \times 10^6$ weights for the Bernstein polynomials, 95$\%$ of which are lesser than 10$^{-8}$ (the absolute tolerance adopted during numerical integration).

We calculate the joint distribution $f(M_p,R_p, M_\star, S)$, which is then conditioned on different quantities to predict the expected planetary mass as a function of planetary radius, stellar mass and insolation flux  --- $f(M_p | R_p, M_\star, S)$. We characterize one source of uncertainties on this conditional distribution using a Monte-Carlo approach where the input sample is perturbed within the measurement uncertainties, and then re-fit to obtain a new set of weights, joint distribution and predictions. Furthermore, since this analysis has been performed in sample space (without accounting for detection biases), we also bootstrap resample (with replacement) the data 100 times to estimate the impact of the small sample size in 4-D when making predictions with the model (the model uncertainty due to the finite sample size is further described in Appendix \ref{sec:sample_size}). For this application, we find that  the variance from bootstrapping the sample exceeds the Monte-Carlo uncertainties.


Then we convert the predicted mass into bulk density\footnote{To avoid confusion between bulk planetary density and the statistical usage of density, i.e., probability density, we use bulk density for the physical quantity (i.e., \gcmcubed) and density for the statistical probabilities.} to consider the change in bulk density for planets of different radii, insolation fluxes and stellar masses (\autoref{fig:bulkdensity_stmass4d} and \autoref{fig:bulkdensity_insol4d}). This example demonstrates an application of this technique.  While we note a few preliminary trends here, we caution against over-interpretation given the heterogeneous nature of the dataset and the complication of inhomogeneous detection completeness\footnote{Such a fit can be performed on a simulated dataset based on the well-characterized \Kepler~ data \citep{hsu_occurrence_2019, he_debiasing_2022} to reduce the impact of survey incompleteness.}. More detailed analyses and the scientific interpretations of the predictions are left to future work. \autoref{fig:bulkdensity_stmass4d} shows that (i) the detected sub-Neptunes ($R_p > 2.5$ \earthradius) have fairly constant bulk densities across insolation; (ii) the detected Earth radius objects tend to have densities higher than Earth. We caution that the trends seen at lower insolation for the 0.7 and 1.0 \solmass~case have a large variance estimated from the bootstraps, which suggests that the predictive power in this region is low due to a small number of data points.

Similarly, \autoref{fig:bulkdensity_insol4d} shows preliminary trends with stellar mass where we see an increase in the bulk densities of the detected super-Earths ($R_p \sim 1.5$ \earthradius) with stellar mass, almost by a factor of two (from 4 \gcmcubed~to 9 \gcmcubed) between 0.3 \solmass~and 1.0 \solmass, whereas this effect is not seen for the gaseous sub-Neptunes. Since the RV semi-amplitude precision has been limited to 1 \ms{} up until recently, this trend could be due to the enhanced RV signatures of these small planets around lower-mass stars.  While the high bulk density for rocky planets around solar-type stars could potentially be at least partially due to a detection bias, this cannot explain the lack of comparable high bulk density super-Earths around the lower-mass M-dwarfs ($< 0.6$ \solmass). This trend is seen across the samples for insolation fluxes 50 \earthinsol~and above.  In contrast, super-Earths around M-dwarfs tend to be much lower in bulk density, potentially indicative of water-worlds \citep[50\% water mass + 50\% silicate mass fraction;][]{zeng_growth_2019, luque_density_2022}, though \cite{rogers_conclusive_2023} suggest that this bulk density could also be explained by volatile rich H/He dominated atmospheres.



\begin{figure}[ht]
\fig{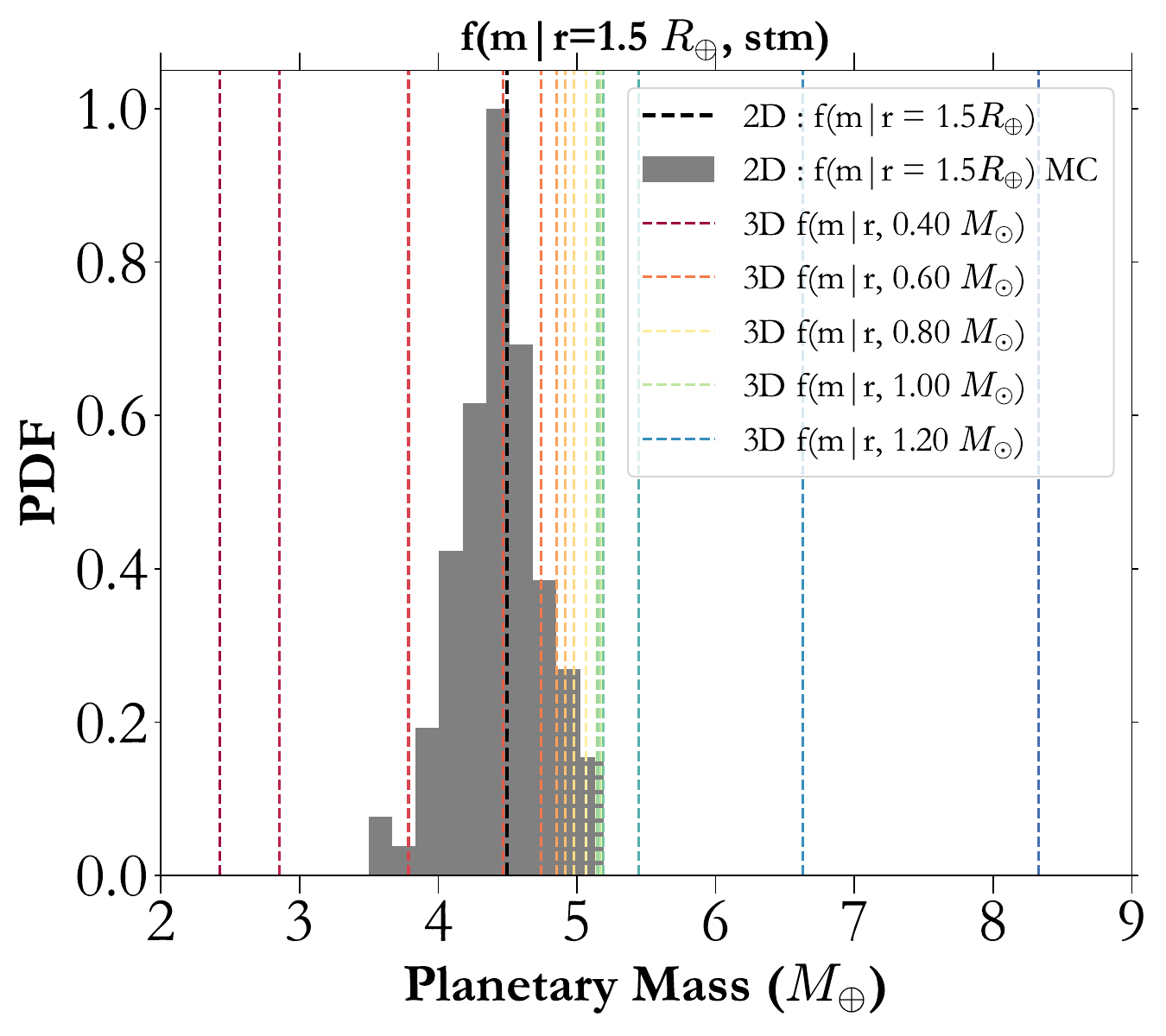}{0.45\textwidth}{}
\fig{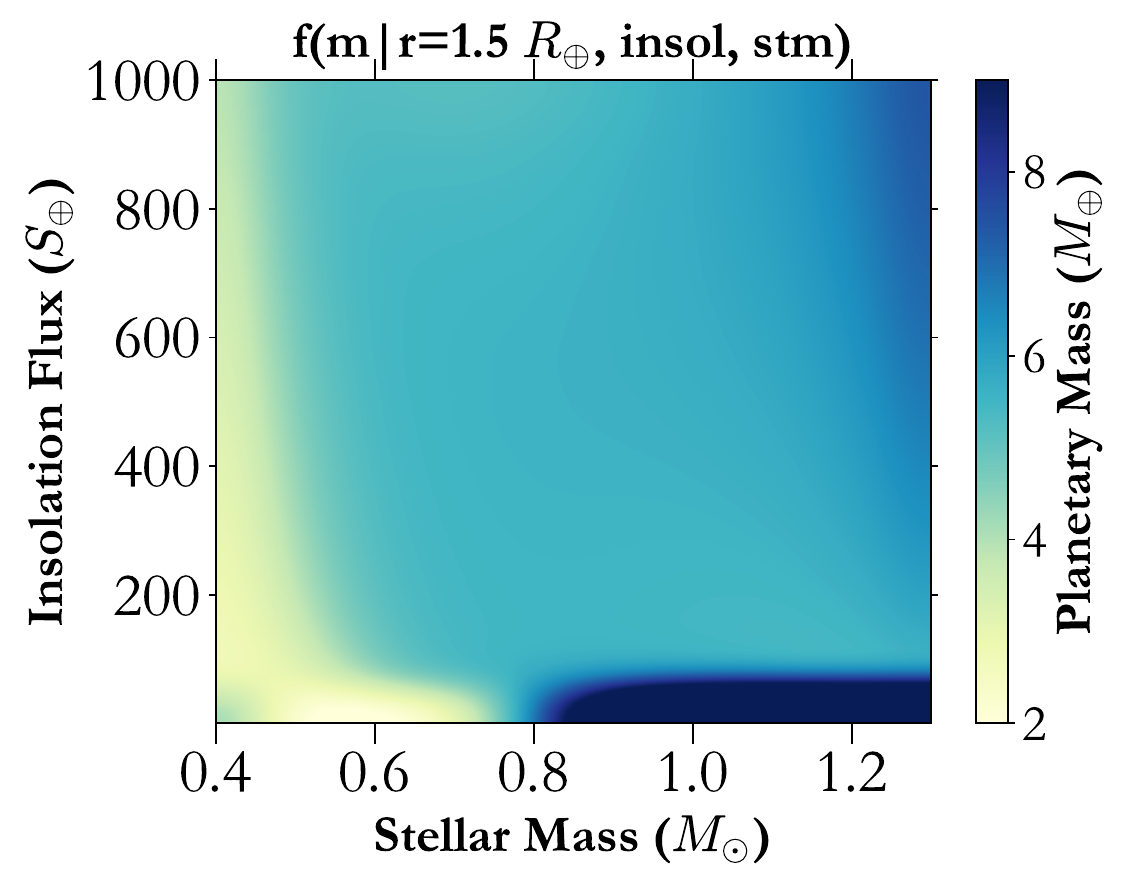}{0.5\textwidth}{}
\caption{\small Comparing Mass-Radius distributions after incorporating additional dimensions for a 1.5 \earthradius planet. \textbf{Top:} The various coloured dashed lines represent the expectation value of $f(m|r,stm)$, whereas the black dashed line represents the prediction from a 2D $f(m|r)$ distribution. The grey histogram shows the distribution of predictions from a Monte-Carlo simulation on the input 2D dataset. \textbf{Bottom:} Predicted planetary mass as a function of stellar mass and insolation flux where the colour bar represents the expectation value for the planetary mass compared to the 2D prediction of $\sim$ 4.5 \earthmass.}\label{fig:MR2d_vs_3d}
\end{figure}

Finally, as a follow-up to the predictive function included with \texttt{MRExo} to predict planetary mass from radius for samples of \Kepler~(FGK hosts) and M-dwarf planets \cite{kanodia_mass-radius_2019}, we include a predictive function based on planetary mass, radius, insolation and stellar mass called \texttt{calculate\_conditional\_distribution()} with the new version of \texttt{MRExo} released with this manuscript. The results for the fit are included on \texttt{Zenodo} along with sample-scripts on \texttt{GitHub}\footnote{\href{https://zenodo.org/record/8222163}{DOI: 10.5281/zenodo.8222163}}.

As TESS is contributing to the sample of planets with measured masses, the range of stellar masses and insolation fluxes covered by the planets is no longer restricted to predominantly short-period objects around Solar-type stars. This is evident in the sample shown in \autoref{fig:MR2d}. The impact of considering these additional dimensions is shown in \autoref{fig:MR2d_vs_3d} with the same dataset and fit presented above, where planetary massses for a 1.5 \earthradius{} planet can change by more than 5x across this parameter space.

\begin{figure}[ht!]
\centering
\includegraphics[scale=0.45,trim={0.4cm 0.2cm 0.2cm 0.2cm},clip]{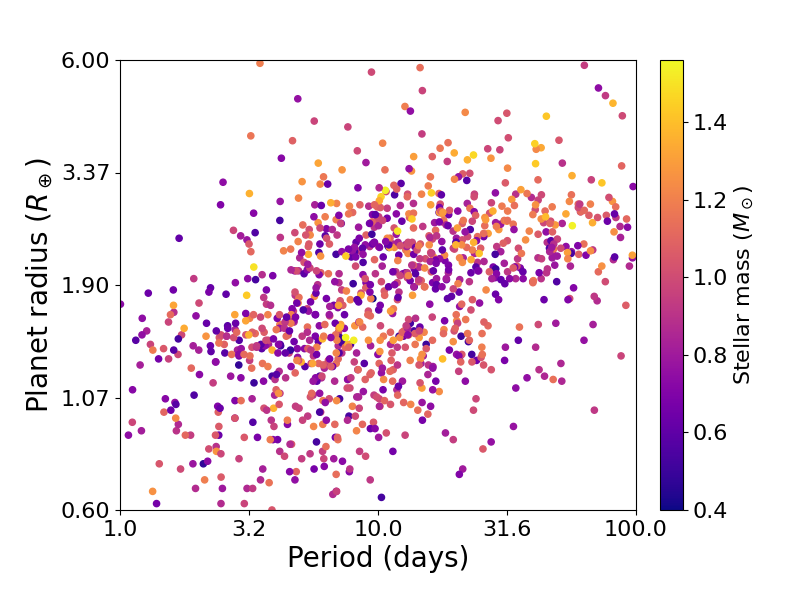}
\caption{The CKS-X sample in plotted in period--planet radius--stellar mass ($P$-$R_p$-$M_\star$), consisting of 1073 planets given our cuts as described in Section \ref{sec:applications:cks-x}. Each point denotes a planet, where the color denotes its host stellar mass.
}
\label{fig:CKS-X_data_period_radius_stmass}
\end{figure}

\subsection{CKS-X data: the Exoplanet Radius Valley} \label{sec:applications:cks-x}

\begin{figure*}[ht]
\centering
\begin{tabular}{cc}
 \includegraphics[scale=0.45,trim={0.4cm 0.2cm 0.2cm 0.2cm},clip]{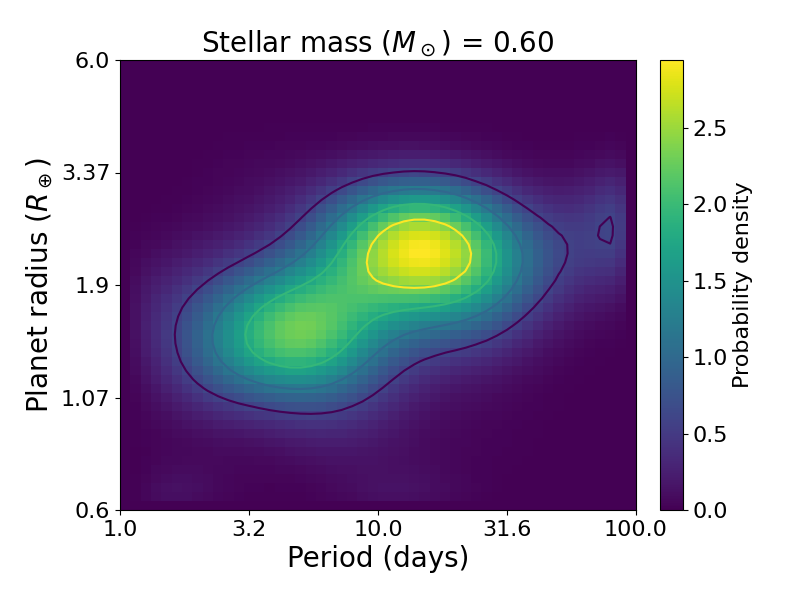} &
 \includegraphics[scale=0.45,trim={0.4cm 0.2cm 0.2cm 0.2cm},clip]{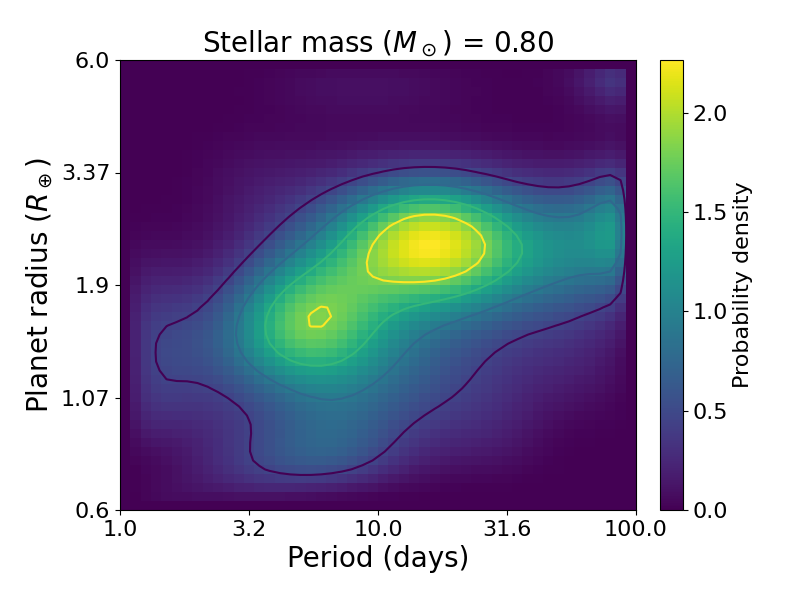} \\
 \includegraphics[scale=0.45,trim={0.4cm 0.2cm 0.2cm 0.2cm},clip]{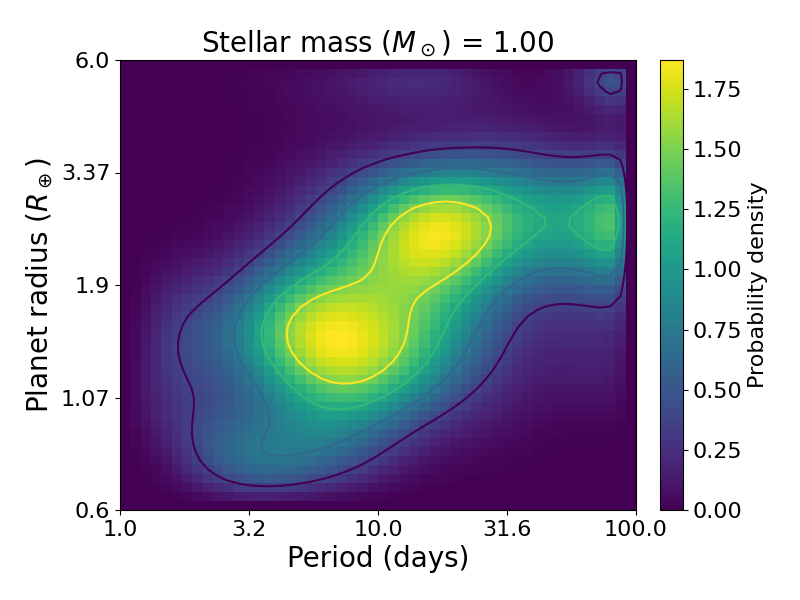} &
 \includegraphics[scale=0.45,trim={0.4cm 0.2cm 0.2cm 0.2cm},clip]{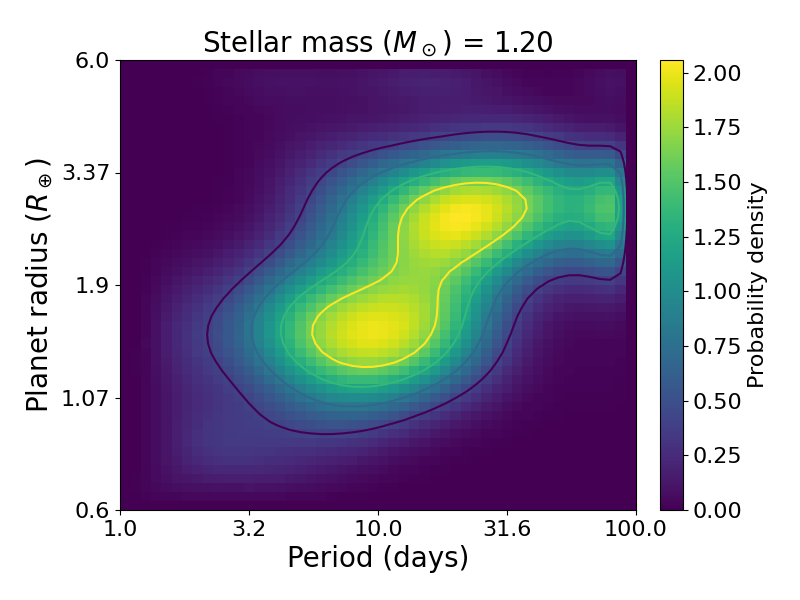} \\
\end{tabular}
\caption{\textbf{Modeling the distribution of planets in 3-D (period--planet radius--stellar mass) using the CKS-X planet sample.} Joint planet radius--period distributions conditioned on various stellar masses (i.e. $f(P,R_p|M_\star)$, where $M_\star = 0.6$, 0.8, 1.0, and $1.2~M_\odot$, as labeled above each panel). The model was fit to the CKS-X sample (1073 planets, as filtered in Section \ref{sec:applications:cks-x}) with a fixed number of degrees in each dimension ($d = 30$, chosen from the cross-validation method). These distributions represent the modeled-observed distributions, as no detection biases were corrected for in any manner. The color-scale in each panel represents the conditional probability density such that each conditional distribution integrates to unity, as computed in $\log{R_p}$-$\log{P}$ space.}
\label{fig:CKS-X_period_radius_cond_stmass}
\end{figure*}

Here, we use  \texttt{MRExo} to apply our model framework to the sample of exoplanets from the California-\Kepler{} Survey \citep[CKS-X; ][]{petigura_california-kepler_2022}. The CKS-X sample is a subset of the \Kepler{} DR25 planet catalog with precise stellar properties as measured from optical spectra obtained on Keck/HIRES \citep{vogt_hires_1994}. This sample builds upon the catalog presented in CKS-I \citep{petigura_california-kepler_2017} by incorporating additional stellar spectra of planet-hosting stars extending down to $\sim 0.4~ M_\odot$, expanding the original sample of spectra for stars in the $\sim 0.8-1.4~M_\odot$ range, for a total of 1246 KOIs orbiting 888 host stars \citep{petigura_california-kepler_2022}. The CKS-X sample thus provides an excellent dataset for modeling the planet period-radius distribution as a function of stellar host properties, as already demonstrated in \citet{petigura_california-kepler_2022}.

While \citet{petigura_california-kepler_2022} characterized the joint $P$-$R_p$ distribution using a series of Gaussian kernel density estimates (KDEs) by dividing the sample into several stellar mass bins, our approach enables the simultaneous fitting of the full 3-D (or even higher-dimensional) distribution using a joint $P$-$R_p$-$M_\star$ distribution with which we can condition on any stellar mass in the range constrained by the data.
In Figure \ref{fig:CKS-X_data_period_radius_stmass}, we plot the CKS-X sample from \citet{petigura_california-kepler_2022} in $P$-$R_p$-$M_\star$ space. We apply a few minor filters and modifications to the CKS-X data before fitting the non-parametric model, by: (1) keeping only planets with orbital periods in the range of $P = [1,100]$ days and radii in the range $R_p = [0.6, 6] R_\oplus$ (which also filters out some objects with spuriously large values), (2) keeping only planets around stars with stellar masses between $M_\star = [0.4, 1.6] M_\odot$ and removing those with no stellar mass uncertainties (`E\_Mstar-iso'$ = 0$), and (3) assuming no uncertainties in the orbital periods.
This results in 1073 remaining planets, for which we fit a model to their joint $P$-$R_p$-$M_\star$ distribution using 30 degrees for each dimension, chosen to be close to the optimal number of degrees from the cross-validation method (we note that the AIC method chooses much fewer degrees, $d \sim 10$). In Figure \ref{fig:CKS-X_period_radius_cond_stmass}, we show the resulting joint $P$-$R_p$ distributions conditioned on various values of stellar mass, $f(P,R_p|M_\star)$. In each panel, the radius valley is clearly visible as the relative dip between two modes of peak probability density.
The detection efficiency decreases for smaller and longer period planets, and this clearly contributes to the observed decrease in density at the smallest sizes and longest period.  However, the detection efficiency varies smoothly and does not have a local maximum that would lead to the local minimum in the $P$-$R_p$ space that could cause the radius valley to be due to select effects.  This is confirmed by other non-parametric population analyses that do model the complex detection efficiency of the \Kepler{} mission \citep{hsu_occurrence_2019, kunimoto_searching_2020, bryson_occurrence_2021}.  We also note that the weights near the boundaries of each dimension can be less reliable. To guard against this, the chosen bounds should be away from regions of scientific interest if feasible. Another possibility is to try joint fits with and without the edge polynomials, and quantify the impact on inferred conditional PDFs using standard distribution comparison metrics. 

While some methods have been recently devised to fit the radius valley using a linear relation \citep[see e.g.,][]{berger_evidence_2023}, their results are sensitive to the exact procedure and we do not attempt to fit a functional form to the exact location of the radius valley in this work. Yet, our non-parametric model provides an avenue for future studies to characterize the radius valley as a function of stellar properties that has at least two advantages: (1) it does not rely on discretizing the data into various bins (e.g. of stellar mass), and (2) the radius valley can be fit to the full (e.g., 3-D) joint distribution that is characterized by a flexible, probabilistic model, instead of fitting to slices of kernel density estimates \citep[KDEs, as in][]{berger_evidence_2023}. From Figure \ref{fig:CKS-X_period_radius_cond_stmass}, we make the qualitative observation that the location of the radius valley (in terms of planet radius) appears to increase with stellar mass, consistent with previous findings with the CKS data \citep{berger_gaia-kepler_2020} and predictions from theoretical models for photoevaporation \citep[e.g.,][]{owen_kepler_2013, wu_mass_2019} and core-powered mass loss \citep{gupta_signatures_2020}.

\subsection{Class II Protoplanetary Disk Dust masses}
\begin{figure}[] 
\centering
\includegraphics[width=0.5\textwidth]{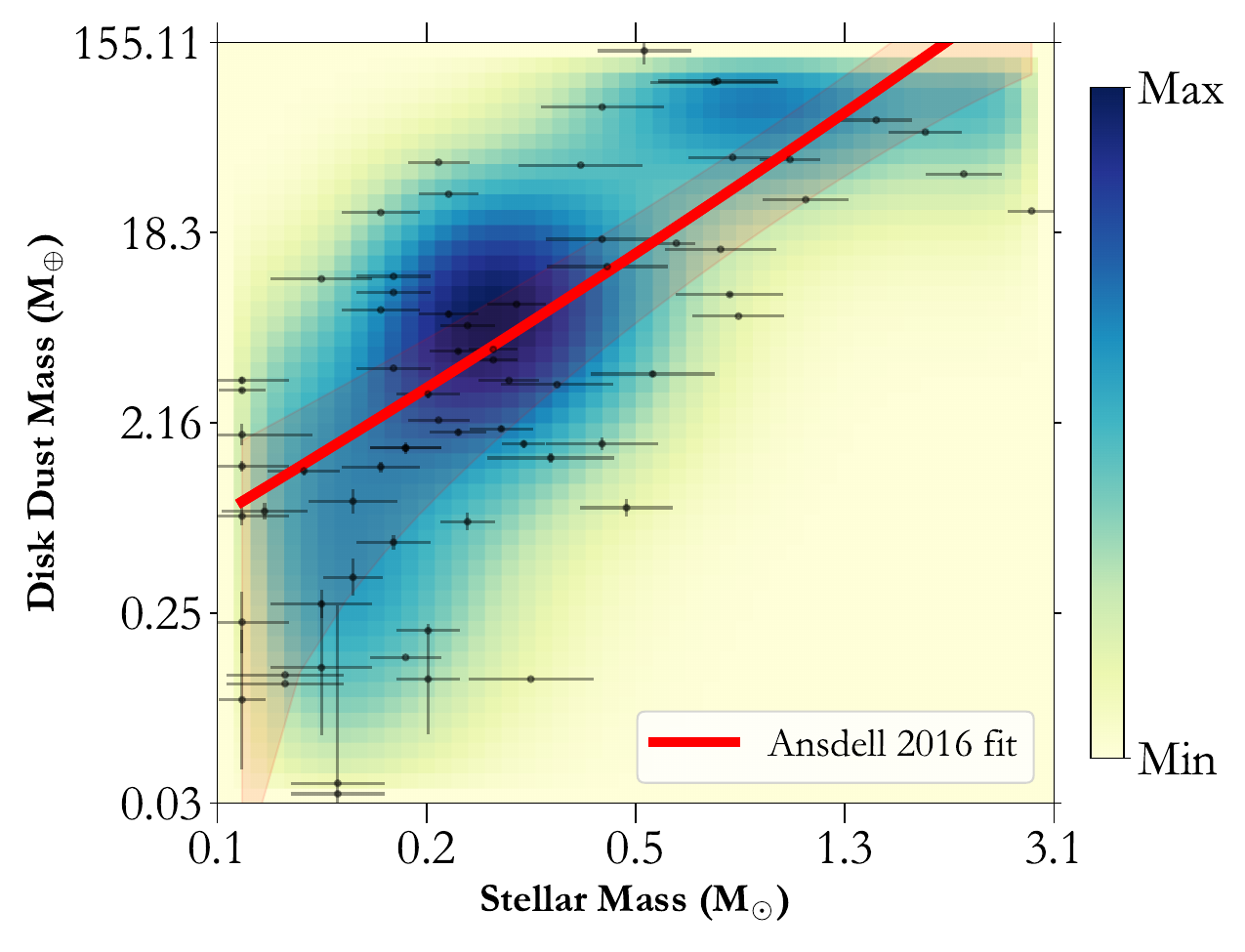}
\caption{Joint distribution of the stellar mass - disk dust mass distribution for 69 disks in the Lupus complex as presented in \cite{ansdell_alma_2016}. Additionally, the red shaded region depicts the power law fit from \cite{ansdell_alma_2016} for reference.} \label{fig:joint_lupus}
\end{figure}

We also use \texttt{MRExo} on a sample of 69 class II protoplanetary disks in the 1 -- 3 Myr Lupus sample based on ALMA observations \citep{ansdell_alma_2016}. Specifically, we perform a joint fit on the stellar mass and  disk dust mass while including the 3-$\sigma$ (99.7 $\%$) upper limits for the latter as a combination of two Gaussian half-normal distributions (described in Section \ref{sec:asymmetric}). Using the cross-validation method, we estimate 15 degrees in each dimension, and calculate the 2-d joint distribution (\autoref{fig:joint_lupus}).

\begin{figure}[] 
\centering
\includegraphics[width=0.5\textwidth]{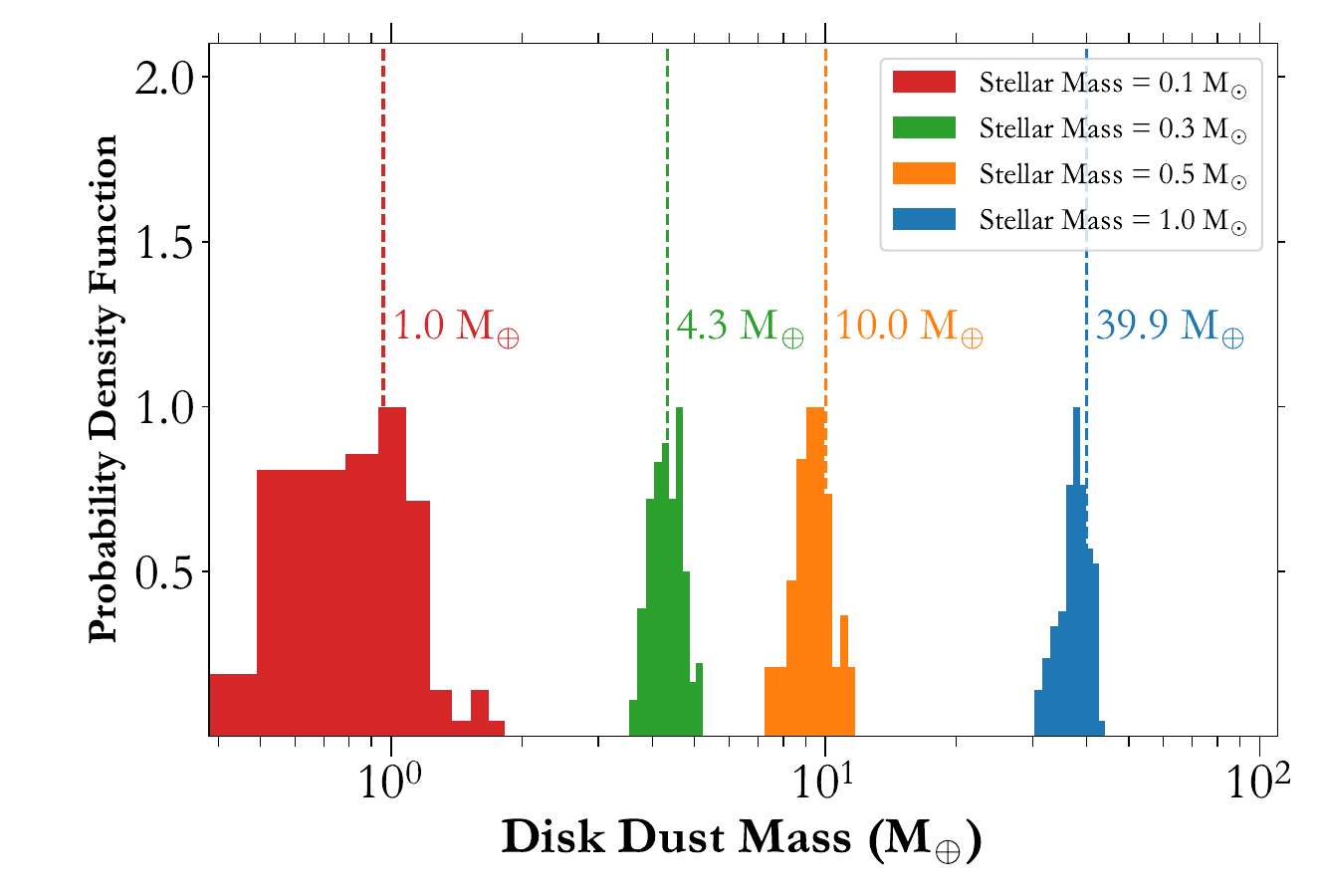}
\caption{Conditional distribution predicting the disk dust mass for different stellar masses.  The dashed lines and the masses are the expected value (\autoref{eq:Expectation}) for each radius, whereas the histogram shows the predictions from the Monte-Carlo simulation. } \label{fig:conditional_diskdust}
\end{figure}

Similar to the approach earlier, we condition the 2-D joint distribution on a few different stellar masses to obtain posteriors for predicted dust masses based on the given sample, along with their Monte-Carlo uncertainties \autoref{fig:conditional_diskdust}, thereby demonstrating the utility of this approach on a different dataset than exoplanet mass-radius.

\section{Conclusion}\label{sec:conclusion}
We build upon the 2-dimensional nonparametric framework utilizing beta density functions as the basis set for density estimation \citep{ning_predicting_2018, kanodia_mass-radius_2019} to perform simultaneous density estimation in up to 4-dimensions. Furthermore, we also modify the existing algorithm to allow measurement upper (and lower) limits to be fit. We discuss the caveats and the degeneracies in log-likelihood space associated with this dimensional expansion, and also run some simulations to demonstrate the utility of the bootstrap and Monte-Carlo methods to explore the impact of the finite sample size and measurement precision of the dataset, respectively, on the inferred predictions. 
We summarize some of the salient features of this framework below:
\begin{itemize}
    \item The non-parametric nature of the framework makes it agnostic to most\footnote{It does assume that the joint density is continuous and smoothly varying, which are desirable properties of a model/well-behaved function. Our implementation also assumes that the density is bounded within the chosen box for the parameter space, since we set the weights along the $n$-dimensional boundary to zero; however, in principle this choice can be relaxed.} assumptions for an intrinsic functional form (e.g., linear or power-law, etc.) and thus also very flexible.
    \item Its probabilistic nature allows one to properly account for both the intrinsic astrophysical spread and measurement uncertainties in the data in a hierarchical framework.
    \item The model treats all dimensions symmetrically, performing a joint fit for the full $n$-dimensional distribution that does not assume that any dimension is dependent on another (e.g., planet mass as a function of radius or vice versa).
    \item The model framework naturally generalizes to higher dimensions ($n \ge$ 2).
\end{itemize}

Motivated primarily by the final point above, we expand the framework by introducing the updates summarized below:
\begin{itemize}
    \item We generalize the nonparametric model to be fit to any number of dimensions (Appendix \ref{sec:joint}, \ref{sec:calc_likelihood}). In practice, this approach is feasible for performing joint fits in up to four dimensions (limited by the available memory for constructing the multi-dimensional arrays).
    \item We switch the optimizer used to calculate the coefficients for each weight to the MM-algorithm which is much more computationally efficient than previous methods (Appendix \ref{sec:optimize_likelihood}).
    \item It can now account for asymmetric measurement uncertainties (i.e, different upper and lower error bars), as well as measurement upper limits, by treating the probability density function as a mixture of two half-normal distributions (Appendix \ref{sec:asymmetric}).
    \item We generalize the framework to allow for different degrees (i.e., varying levels of resolution or complexity) in each dimension (Appendix \ref{sec:degree_selection}).
    \item We provide two different methods for choosing the number of degrees by maximizing the log-likelihood and finding the optimum number of degrees: (i) the cross-validation (CV) method, and (ii) the AIC method. We find that the AIC method tends to return a lower number of degrees than cross-validation in the example applications considered (Appendix \ref{sec:degree_selection}).
    \item To quantify the model uncertainties due to the reduced density of samples in higher dimensional parameter spaces (i.e., the finite sample size), we include a bootstrap sampling algorithm, which can be used to quantify the variance in prediction outcomes due to this effect (Appendix \ref{sec:sample_size}).
    \item We also include the possibility of performing Monte-Carlo sampling on the input dataset to quantify the impact of the measurement uncertainties on the predictions.
\end{itemize}

Finally, we combine these statistical techniques and explore three case studies to showcase the applicability of this methodology. 

\begin{enumerate}
    \item We perform a 4-dimensional fit to a sample of small planets ($R_p < 4$ \earthradius) with mass measurements in the joint distribution of planetary mass, radius, insolation and stellar mass. The model hints at trends in bulk density with insolation for super-Earths and Neptunes. We also see hints in the sample that 1.5 \earthradius{} super-Earths tend to be lower in bulk density around M-dwarfs ($M_\star < 0.6$ \solmass) than FGK host stars. The absence of the higher bulk density super-Earths cannot be a detection bias, and reinforces previous studies of water-world super-Earths (or H/He rich sub-Neptunes) around M-dwarfs \citep{luque_density_2022, rogers_conclusive_2023}. 
    \item We perform a 3-dimensional fit to the CKS-X sample in terms of the planetary radius, orbital period, and stellar mass. This example demonstrates that our nonparametric model can clearly capture the observed radius valley, as well as its dependence on host stellar mass without discretizing the sample into various bins (as was done in previous studies). We also use this example to showcase the utility of bootstrap resampling in masking out the regions in which the mean joint density is poorly constrained by the data (Appendix \ref{sec:sample_size}).
    \item We perform a 2-dimensional fit to a sample of protoplanetary disks in terms of their dust masses and host stellar masses, which offers more flexibility than the simple power-law fits used in previous studies. We demonstrate that this approach allows us to predict disk properties (including the Monte-Carlo uncertainties) for different host-stellar masses while incorporating measurement upper limits.
\end{enumerate}

Alongside this manuscript, we also release the updated version of our free open-source \texttt{python} package -- \texttt{MRExo} -- which allows users to perform their own exploration of different datasets in multi-dimensional space to tease out trends as well as use as a predictive tool for inferences.

\section{Acknowledgements}


We thank Suvrath Mahadevan, Johanna Teske, Gudmundur Stefansson, Anjali Piette and Peter Gao for helpful discussions and feedback regarding this manuscript. SK acknowledges Peter Gao for help with computing resources to perform some of the analysis presented in this manuscript.

The Pennsylvania State University campuses are located on the original homelands of the Erie, Haudenosaunee (Seneca, Cayuga, Onondaga, Oneida, Mohawk, and Tuscarora), Lenape (Delaware Nation, Delaware Tribe, Stockbridge-Munsee), Shawnee (Absentee, Eastern, and Oklahoma), Susquehannock, and Wahzhazhe (Osage) Nations.  As a land grant institution, we acknowledge and honor the traditional caretakers of these lands and strive to understand and model their responsible stewardship. We also acknowledge the longer history of these lands and our place in that history.

Computations for this research were performed on the Pennsylvania State University’s Institute for Computational and Data Sciences Advanced CyberInfrastructure (ICDS-ACI).  This content is solely the responsibility of the authors and does not necessarily represent the views of the Institute for Computational and Data Sciences.

The Center for Exoplanets and Habitable Worlds is supported by the Pennsylvania State University, the Eberly College of Science, and the Pennsylvania Space Grant Consortium. 

This research made use of the (i) NASA Exoplanet Archive, which is operated by Caltech, under contract with NASA under the Exoplanet Exploration Program, (ii) SIMBAD database, operated at CDS, Strasbourg, France, and (iii) NASA's Astrophysics Data System Bibliographic Services.

This research has made use of the SIMBAD database, operated at CDS, Strasbourg, France, 
and NASA's Astrophysics Data System Bibliographic Services.

\software{
\texttt{astropy} \citep{robitaille_astropy_2013, astropy_collaboration_astropy_2018},
\texttt{ipython} \citep{perez_ipython_2007},
\texttt{matplotlib} \citep{hunter_matplotlib_2007},
\texttt{MRExo} \citep[][and this work]{kanodia_mass-radius_2019},
\texttt{numpy} \citep{oliphant_numpy_2006},
\texttt{pandas} \citep{mckinney_data_2010},
\texttt{scipy} \citep{oliphant_python_2007, virtanen_scipy_2020},
}

\appendix

\section{Generalizing to 2+ dimensions}\label{sec:generalizing}
\subsection{Joint Distribution}\label{sec:joint}

Generalizing Equation 7 from \cite{ning_predicting_2018} for the joint distribution from 2 to $n$ dimensions we have, 

\begin{align}\label{eqn:joint}
    & f(x_1,...,x_n|\bm{w}, d^{(1)},...,d^{(n)}) \notag \\ &\quad = \sum_{\tau_{1}=1}^{d^{(1)}}...\sum_{\tau_{n}=1}^{d^{(n)}} w_{\tau_{1}...\tau_{n}} \frac{\beta_{\tau_{1} d^{(1)}}\left(\frac{x_{1} - \underline{X_1}}{\overline{X_1} - \underline{X_1}} \right)}{\overline{X_1} - \underline{X_1}}...\frac{\beta_{\tau_{n} d^{(n)}}\left(\frac{x_{n} - \underline{X_n}}{\overline{X_n} - \underline{X_n}} \right)}{\overline{X_n} - \underline{X_n}}
\end{align}

where, 

\begin{itemize}
\item $t$ iterates through each dimension and $t \in$ \{1,.., n\}
\item $d^{(t)}$ is the number of degrees in each dimension. 
\item $\tau_t$ iterates through $d^{(t)}$ in dimension $t$.  Earlier denoted using $k$, $l$ in \cite{ning_predicting_2018}.
\item $w_{\tau_{1}...\tau_{n}}$ is an element in the $n$-dimensional matrix of weights $\bm{w}$.
\item $x_t$ is the continuous variable used to sample dimension $t$ of sample size $N$
\item $\overline{X_t}$ and $\underline{X_t}$ are the upper and lower bounds for dimension $t$.
\item $\beta_{\tau_{t} d^{(t)}}$ is the beta distribution function, with one of the shape parameters being $\tau_t$ and the other $d^{(t)}$, and the continuous variable $x_t$ is normalized by the upper and lower bounds.
\end{itemize}

\subsection{Calculating Likelihood}\label{sec:calc_likelihood}
There are two main unknown parameters in this model, the matrix of weights $\bm{w}$, and the choice of degrees for each dimension. To estimate these we continue to expand on the formalism from \cite{ning_predicting_2018} and define a likelihood function $\mathcal{L}$ (similar to their Equation 8),

\begin{align}
    & \mathcal{L}(\bm{w}, d^{(1)},...,d^{(n)} ~|~ \bm{X}_1^{obs},...,\bm{X}_n^{obs}, \bm{\sigma}_1^{obs},..., \bm{\sigma}_n^{obs}) \nonumber \\  
    &\quad = \int^{\overline{X_1}}_{\underline{X_1}} ... \int^{\overline{X_n}}_{\underline{X_n}} f(\bm{X}_1^{obs},...,\bm{X}_n^{obs}, x_1, ..., x_n |  \bm{w}, d^{(1)},...,d^{(n)}, \bm{\sigma}_1^{obs},..., \bm{\sigma}_n^{obs}) ~\textrm{d} x_{1} ~...~ \textrm{d} x_{n} \\
    &\quad = \prod_{i=1}^{N} \int^{\overline{X_1}}_{\underline{X_1}} ... \int^{\overline{X_n}}_{\underline{X_n}}  f(X_{1,i}^{obs}|x_1, \sigma_{1,i}^{obs}) ~...~ f(X_{n,i}^{obs}|x_n, \sigma_{n,i}^{obs}) \times f(x_1,...,x_n|\bm{w}, d^{(1)},...,d^{(n)}) ~\rm{d} \textit{x$_{1}$} ~...~ \rm{d} \textit{x$_{n}$}
\end{align}

where 

\begin{itemize}
    \item $i$ iterates through each observed point.  $i \in$ \{1,2,..,N\}
    \item $X_{t,i}^{obs}$ is measured quantity $i$ in dimension $t$, drawn from $\bm{X}_t^{obs}$.
    \item $\sigma_{t,i}^{obs}$ is the uncertainty on the measured quantity $i$ in dimension $t$, drawn from $\bm{\sigma}_t^{obs}$
\end{itemize}

Here the measured quantity is expressed as,

\begin{equation}\label{eqn:NormalPDF}
    f(X_{t,i}^{obs}|x_t, \sigma_{t,i}^{obs}) = \mathcal{N}\left(\frac{X_{t,i}^{obs} - x_{t}}{\sigma_{X^{obs}_{t,i}}} \right),
\end{equation}

where $\mathcal{N}$ is the standard normal distribution.
Therefore the likelihood function $\mathcal{L}$ entails the convolution of the measured probability distribution (which is assumed to be normal) with the beta distribution (from the joint distribution \autoref{eqn:joint}). Then,

\begin{dmath}
    \mathcal{L} =  \prod_{i=1}^{N} \sum_{\tau_{1}=1}^{d^{(1)}}...\sum_{\tau_{n}=1}^{d^{(n)}} w_{\tau_{1}...\tau_{n}} \int^{\overline{X_1}}_{\underline{X_1}} ... \int^{\overline{X_n}}_{\underline{X_n}} \\ \frac{1}{\sigma_{X^{obs}_{1,i}}} \frac{\beta_{\tau_{1} d^{(1)}}\left(\frac{x_{1} - \underline{X_1}}{\overline{X_1} - \underline{X_1}} \right)}{\overline{X_1} - \underline{X_1}} \mathcal{N}\left(\frac{X_{1,i}^{obs} - x_{1}}{\sigma_{X^{obs}_{1,i}}} \right) \textrm{d} x_{1} ~...~
     \frac{1}{\sigma_{X^{obs}_{n,i}}}\frac{\beta_{\tau_{n} d^{(n)}}\left(\frac{x_{n} - \underline{X_n}}{\overline{X_n} - \underline{X_n}} \right)}{\overline{X_n} - \underline{X_n}} \mathcal{N}\left(\frac{X_{n,i}^{obs} - x_{n}}{\sigma_{X^{obs}_{n,i}}} \right) \textrm{d} x_{n}
\end{dmath}

for brevity we introduce $\mathcal{P}_t(\tau_t, i)$, which is essentially the convolved probability contribution for each measurement,
\begin{align}
    &\mathcal{L} = \prod_{i=1}^{N} \sum_{\tau_{1}=1}^{d^{(1)}}...\sum_{\tau_{n}=1}^{d^{(n)}} w_{\tau_{1}...\tau_{n}} \mathcal{P}_1(\tau_1, i) ... \mathcal{P}_n(\tau_n, i)
    \intertext{where $\mathcal{P}_t(\tau_t, i)$  cycles through $N$ observations with the iterator $i$, and is defined below}
     &\mathcal{P}_t(\tau_t, i) =  \int^{\overline{X_t}}_{\underline{X_t}} \frac{1}{\sigma_{X^{obs}_{t,i}}}\frac{\beta_{\tau_{t} d^{(t)}}\left(\frac{x_{t} - \underline{X_t}}{\overline{X_t} - \underline{X_t}} \right)}{\overline{X_t} - \underline{X_t}} \mathcal{N}\left(\frac{X_{t,i}^{obs} - x_{t}}{\sigma_{X^{obs}_{t,i}}} \right) \textrm{d} x_{t} \label{eqn:IndvPDF}
     \intertext{Since $\mathcal{P}_t(\tau_t, i)$ is essentially a constant which can be obtained by numerical integration, following \cite{ning_predicting_2018} we combine these constants into $\bm{c}$, a 2-D matrix   ($m \times N$) where $m$ = $\prod_t^n d^{(t)}$. For example for a 2-D sample of size 20 ($N$ = 20) with degrees $d_1 = 10, d_2 = 12$, $\bm{c}$ would have dimensions 120 $\times$ 20. Then,}
      &c_{(\tau_1, \tau_2,..,\tau_n),i} = \prod_{t}^{n} \mathcal{P}_t(\tau_t, i) \\
    \intertext{Equivalent to equation 9 from \cite{ning_predicting_2018}, the likelihood can then be expressed as the product of this $\bm{c}$ and the weights $\bm{w}$. Here we note that while multiplying with $\bm{c}$ we flatten $\bm{w}$ such that it is a 1-d array of length $m$, where $\sum_j^m w_j = 1$.}
    &\textrm{log}~\mathcal{L} = \sum_{i=1}^n \textrm{log}~ (\bm{c}_i^T \bm{w}) = \sum_{i=1}^n \textrm{log}~ (\sum_{j=1}^m c_{ij}w_j)\label{eq:loglikeoptimize}
\end{align}

While $\bm{c}$ can be computed by numerical integration for a input sample set, we use the MM (EM) algorithm to maximize the log-likelihood in a computationally efficient manner, which is discussed in the next section.

\subsection{Maximizing Likelihood using MM Algorithm}\label{sec:optimize_likelihood}
We also modify the method followed to optimize for the weights of the Bernstein polynomials. \cite{ning_predicting_2018} used the inbuilt $\texttt{R}$ non-linear optimizer \texttt{Rsolnp}, and \cite{kanodia_mass-radius_2019} used the Sequential Least Squares optimization routine in \texttt{scipy} --- \texttt{fmin\_slsqp}. Due to convexity of the function  in \autoref{eq:loglikeoptimize}, we adopt the ``Majorize-Minimization" (MM) prescription to construct an optimization routine\footnote{See \cite{lange_legacy_2022} for a review of the MM algorithm}, where we maximize the log-likelihood through the following $r$ iterations in optimization, after initializing $\bm{w}$ as:

\begin{align}
    \bm{w}^{(0)} &= \left( \frac{1}{m}, \frac{1}{m}, ..., \frac{1}{m} \right) \\
    \intertext{then,}
    w^{(r)}_j &= \frac{1}{n} \sum_{i=1}^n \frac{c_{ij} w_j^{(r-1)}}{\sum_{k=1}^m c_{ik} w_k^{(r-1)}} ~\forall~r \in (1,2,..,)
\end{align}

where we stop iterating when $| \textrm{log}~\mathcal{L}^{(r )} -  \textrm{log}~\mathcal{L}^{(r-1)}| \le \epsilon~|\textrm{log}~\mathcal{L}^{(r-1)}|$  where $\epsilon$ = 10$^{-3}$. This typically converges in $<$ 20 iterations, and is much faster than the black-box solvers available in $\texttt{R}$ or $\texttt{python}$. When benchmarked on the 127 planet sample from \cite{ning_predicting_2018} we find the log-likelihood to converge in 0.06 seconds compared to a few hours using \texttt{fmin\_slsqp}.

\section{Asymmetric Errorbars}\label{sec:asymmetric}
Astronomical measurements are rarely associated with Normal (Gaussian) uncertainties. For example, orbital eccentricities need to be positive and finite, which can bias their estimates or posteriors \citep{lucy_spectroscopic_1971}. Often due to instrumental limitations or astrophysical confounding factors, observations are not precise enough to obtain statistically significant (say at 3$\sigma$ or 5$\sigma$) measurements, in which case measurement upper limits are reported at some confidence level (95\% or 99.7\%). This is particularly common for planetary mass measurements where 3$\sigma$ (99.7\%) mass upper limits are often used\footnote{See \cite{plavchan_radial_2015} and \cite{figueira_deriving_2018} for a review of planetary mass measurements using the RV technique.}, or in protoplanetary disk flux measurements, where for faint disks, the flux upper limits can be reported\footnote{See \cite{miotello_setting_2022} for a review on the measurements of fundamental protoplanetary disk properties.}.

To incorporate these measurements into our framework, we account for the possibility of asymmetric measurement errors (with $\sigma_u$ and $\sigma_l$) for each data point ($\bm{X}^{+\bm{\sigma}_u}_{-\bm{\sigma}_l}$) in the sample by modifying \autoref{eqn:NormalPDF} as:

\begin{equation}\label{eqn:HalfNormal}
    f(X_{t,i}^{obs}|x_t, \sigma_{t,i, u}^{obs}, \sigma_{t,i, l}^{obs}) = \mathcal{N}_{+}\left(\frac{X_{t,i}^{obs} - x_{t}}{\sigma_{X^{obs}_{t,i, u}}} \right) + \mathcal{N}_{-}\left(\frac{X_{t,i}^{obs} - x_{t}}{\sigma_{X^{obs}_{t,i, l}}} \right),
\end{equation}

where $\mathcal{N}_{+}$ ($\mathcal{N}_{-}$) is the upper (lower) standard half-normal distribution.
Finally the convolved probability for each measurement (\autoref{eqn:IndvPDF}) becomes

\begin{equation}\label{eqn:HalfNormalPDF}
    \mathcal{P}_t(\tau_t, i) =  \int^{X_{t,i}^{obs}}_{\underline{X_t}} \frac{1}{\sigma_{X^{obs}_{t,i,l}}}\frac{\beta_{\tau_{t} d^{(t)}}\left(\frac{x_{t} - \underline{X_t}}{\overline{X_t} - \underline{X_t}} \right)}{\overline{X_t} - \underline{X_t}} \mathcal{N}_{-}\left(\frac{X_{t,i}^{obs} - x_{t}}{\sigma_{X^{obs}_{t,i,l}}} \right) \textrm{d} x_{t} + 
    \int^{\overline{X_t}}_{X_{t,i}^{obs}} \frac{1}{\sigma_{X^{obs}_{t,i,u}}}\frac{\beta_{\tau_{t} d^{(t)}}\left(\frac{x_{t} - \underline{X_t}}{\overline{X_t} - \underline{X_t}} \right)}{\overline{X_t} - \underline{X_t}} \mathcal{N}_{+}\left(\frac{X_{t,i}^{obs} - x_{t}}{\sigma_{X^{obs}_{t,i,u}}} \right) \textrm{d} x_{t}
\end{equation}

For example, for typical mass upper limits only the 2$\sigma$ (95\%) or 3$\sigma$ (99.7\%) upper limit is reported, and not the median value. If we have a measurement with a 2$\sigma$ (95\%) upper limit of 10 \earthmass, then we assume $X^{obs} \equiv \underline{X}$ , such that the lower half-normal PDF $\mathcal{N}_{-} \to 0$ in \autoref{eqn:HalfNormalPDF}, and we estimate $\sigma_u$ such that the upper half-normal PDF integrates to 97.5\% (instead of 95\% since it is a half-normal PDF reproducing the upper limit) at 10 \earthmass. We note the caveat that posteriors in orbital parameters (such as eccentricity and $\omega$) are often non-Gaussian, and thus recommend authors to also report posteriors for the variables Monte-Carlo sampled such as $e$cos$\omega$, $e$sin$\omega$, etc. that are more likely to be Gaussian \citep{lucy_spectroscopic_1971, fulton_radvel_2018}.

\section{Degree Selection}\label{sec:degree_selection}
\begin{figure*}[ht]
\centering
\begin{tabular}{cc}
 \includegraphics[width=8cm]{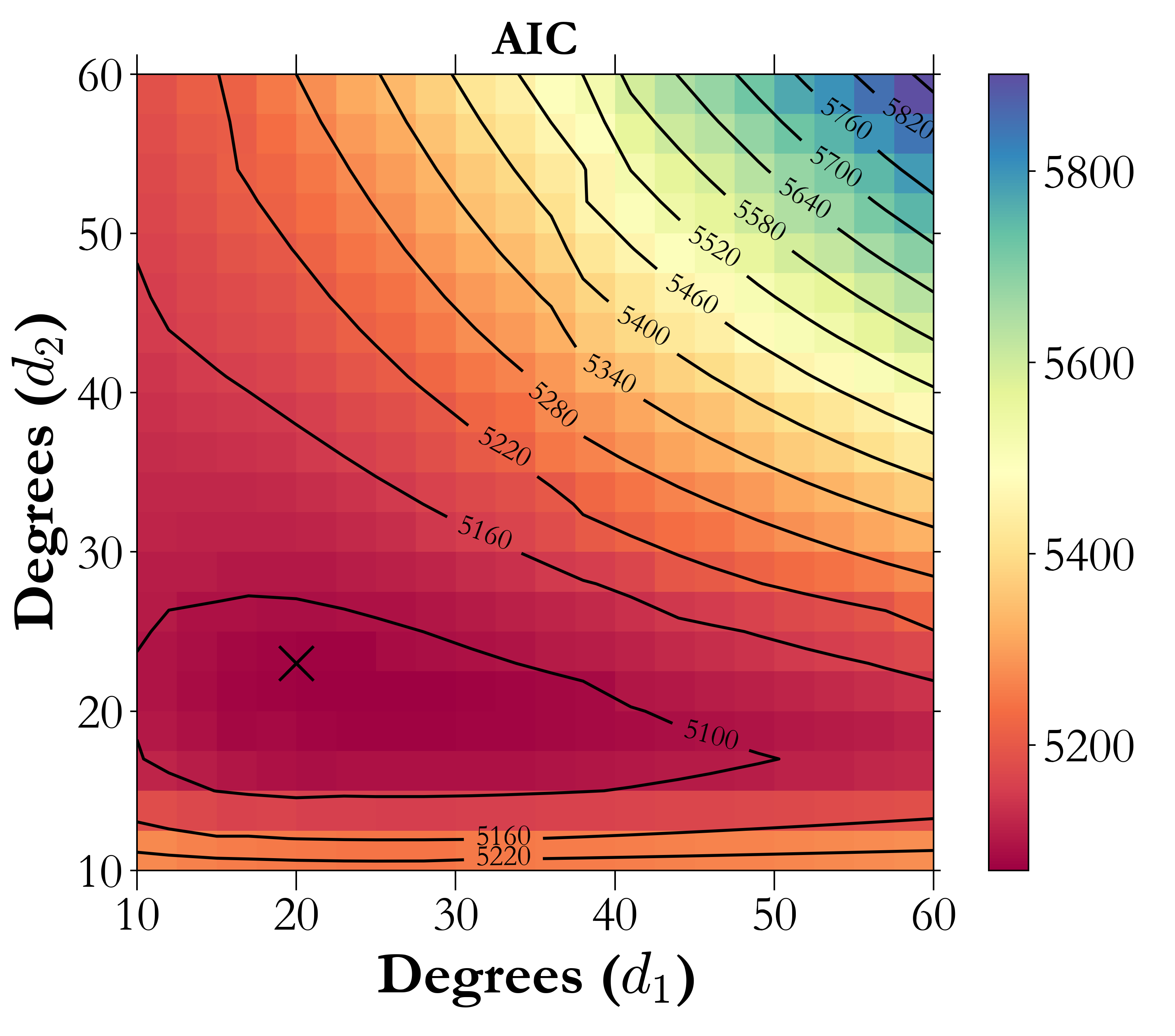} &
 \includegraphics[width=8cm]{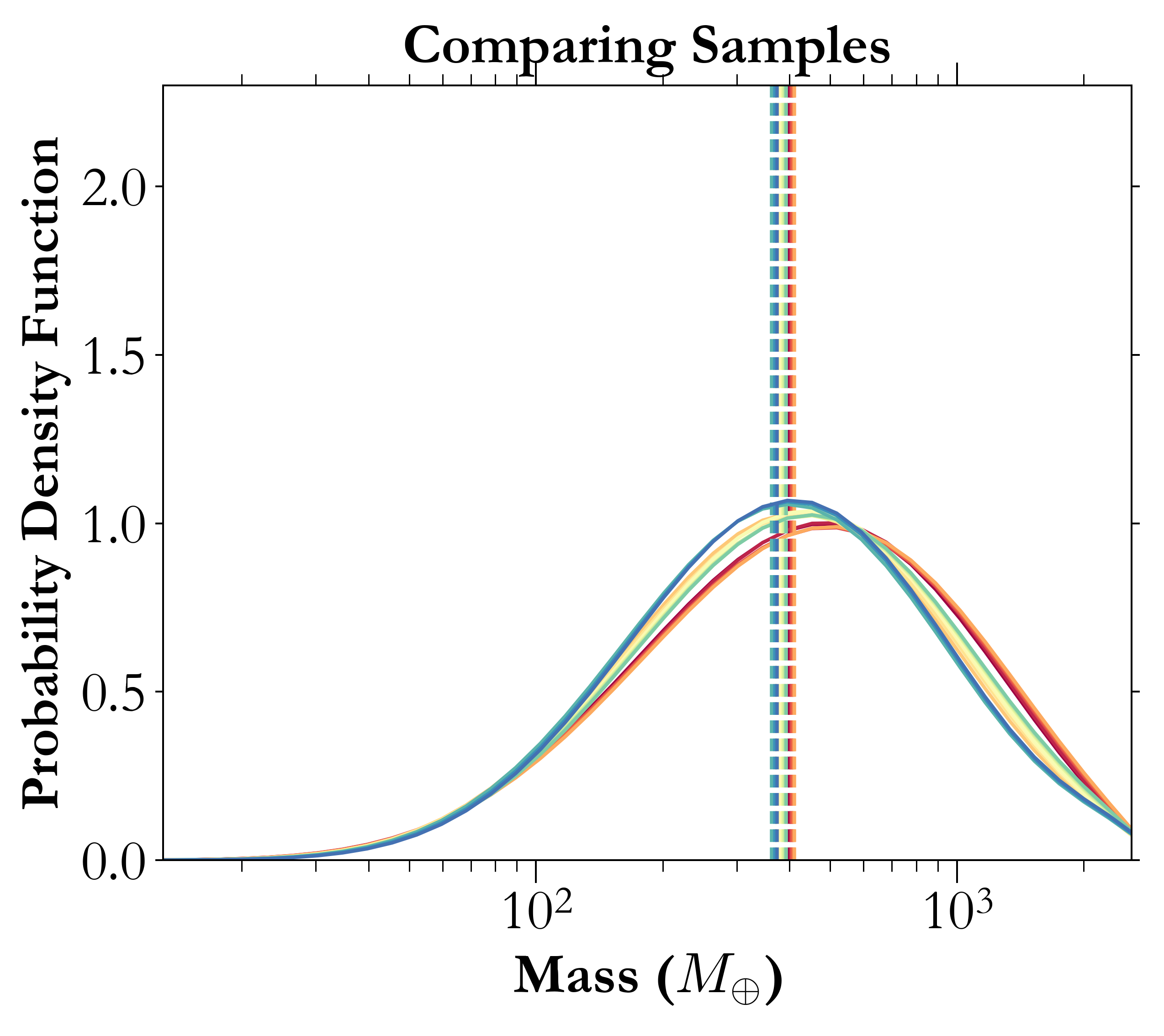} \\
\end{tabular}
\caption{\textbf{a)} Showing a 2-D grid of AIC for the MR dataset shown in (\autoref{fig:joint}), along with contours of similar AIC. In this case, the optimized degrees are roughly equal (20,23) and are marked with an `X'. \textbf{b)} The change in conditional distribution $f(m|r = 12$ \earthradius) as the degree choices are sampled from within the innermost AIC contour \textbf{(a)}. We conclude that for such a dataset with a large amount of scatter, the conditional distribution is not too sensitive to the exact choice of degrees.}   \label{fig:AIC_asymm}
\end{figure*}

The degrees represent the shape of the beta distribution. Modifying previous versions of the algorithm from \cite{ning_predicting_2018} and \cite{kanodia_mass-radius_2019}, we allow for the possibility of different degrees in each dimensions, which should allow the user to use \texttt{MRExo} for density estimation across parameters with different levels of complexity. By default, we sample 10 degree candidates for each dimension and then use the AIC or cross-validation method to pick the optimum degree combination $d^{(1)},...,d^{(n)}$, where the latter is described by \cite{ning_predicting_2018}. The AIC metric is given by $2k - \ln(\mathcal{L})$, where $\ln(\mathcal{L})$ is the log-likelihood described in Section \ref{sec:optimize_likelihood}, and $k$ is the effective number of weights or the effective sample size, which we compute using Design Effect \citep{kish1965survey}, and is given by $k$ = $1/\sum w_i^2$. We show a sample 2-D grid of AIC in \autoref{fig:AIC_asymm}a. To investigate the impact on the conditional distribution, of degree selection within the final contour, i.e., where the AIC values are roughly similar, we fit a range of models for the same dataset with degree choices sampled from the innermost (lowest AIC) contour. Based on \autoref{fig:AIC_asymm}b, we conclude that the conditional distribution is not very sensitive to the exact choice of degrees when the input dataset has large measurement errors or intrinsic scatter as seen in \autoref{fig:joint}. Aside from the AIC method, we also extend the k-fold cross-validation approach from \cite{ning_predicting_2018} to higher dimensions. 

For \texttt{MRExo} users, we suggest starting with a simple optimization with an equal number of degrees for quick checks (sampling through 10 degree candidates instead of $10^n$ for $n$-dimensions). Subsequently, one can perform a more detailed analysis by exploring a full grid of degree candidates which allows for different degrees in each dimension. This has been implemented using a boolean \texttt{SymmetricDegreePerDimension} function call, which then utilizes multiple cores to explore each degree choice with parallel computing implemented through the \texttt{multiprocessing} module in Python.

\section{Understanding the effect of sample size in n-dimensions}\label{sec:sample_size}

\begin{figure*}[ht]
\centering
\begin{tabular}{cc}
 \includegraphics[scale=0.45,trim={0.4cm 0.2cm 0.2cm 0.2cm},clip]{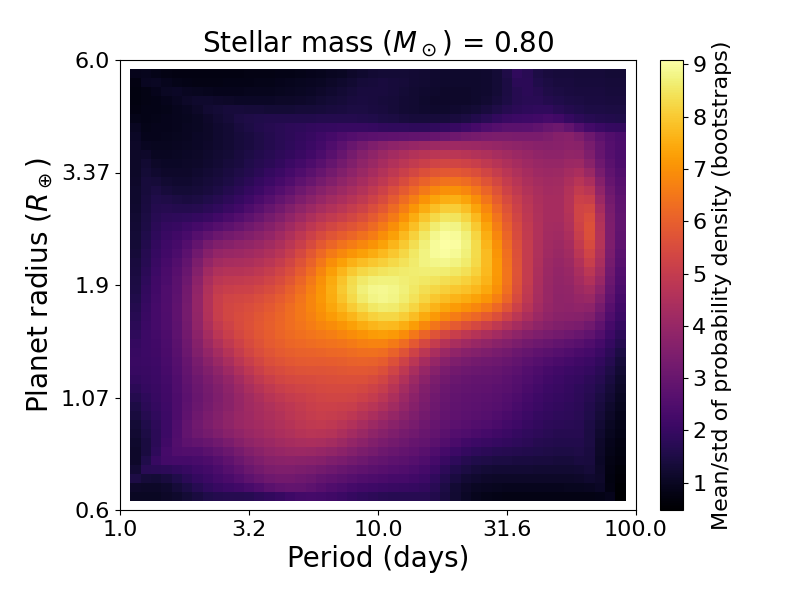} &
 \includegraphics[scale=0.45,trim={0.4cm 0.2cm 0.2cm 0.2cm},clip]{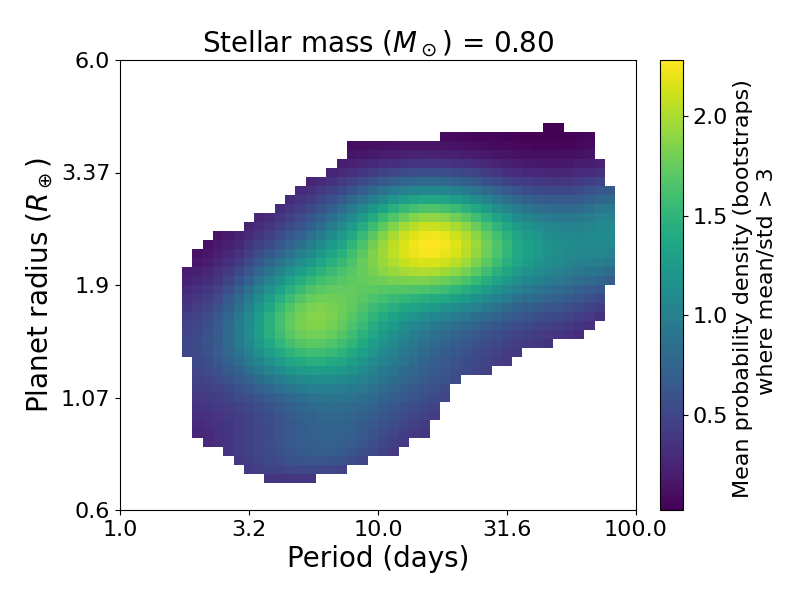} \\
\end{tabular}
\caption{The effect of finite sample size on the joint $P$-$R_p$ distribution conditioned on a given stellar mass, $f(P,R_p|M_\star = 0.80~M_\odot)$, from 100 bootstrap resamplings of the CKS-X dataset (described in Section \ref{sec:applications:cks-x}). \textbf{Left:} the mean divided by the standard deviation of the joint probability densities, $\mu_{f(P,R_p|M_\star)}/\sigma_{f(P,R_p|M_\star)}$, over the bootstraps. Higher values denote regions where the model is more robust due to a greater density of data points. \textbf{Right:} the mean joint probability density of the bootstraps, with the regions $\mu_{f(P,R_p|M_\star)}/\sigma_{f(P,R_p|M_\star)} < 3$ masked out.}
\label{fig:CKS-X_period_radius_cond_stmass_bootstraps}
\end{figure*}

In addition to the uncertainty in the model arising from the measurement errors of the data points, there is also uncertainty due to the finite sample size. This is a byproduct of performing the analysis based on a finite sample of points from the target distribution rather than from the target distribution itself.. For example, there can be significant variance in the \textit{mean} prediction in a region of parameter space where there is a relatively low density of data points, even when the data points in that region are known precisely (i.e. have small measurement errors). This is especially problematic when fitting the model in higher dimensions, since the volume of parameter space grows so rapidly that it is often impractical to collect enough data to maintain a high density of samples. To account for this source of uncertainty in which only one or a few data points strongly dominate the model behavior in some regions, we use bootstrap resampling of the data (with replacement).\footnote{This has previously been done in e.g. \cite{ning_predicting_2018} to quantify the confidence intervals of the mean prediction separately from the predictive intervals around the mean (which capture the intrinsic spread in the data).}

In Figure \ref{fig:CKS-X_period_radius_cond_stmass_bootstraps}, we show the results of the model fits to 100 bootstrap resamplings of the data for the CKS-X dataset (from Section \ref{sec:applications:cks-x}, with degrees set to 30), in terms of their joint $P$-$R_p$ distributions conditioned on a given stellar mass, $f(P,R_p|M_\star = 0.80~M_\odot)$. The left panel shows the mean joint probability density ($\mu_{f(P,R_p|M_\star)}$) divided by the standard deviation of the joint probability densities ($\sigma_{f(P,R_p|M_\star)}$), over the 100 bootstraps. The reason we use the ratio $\mu_{f(P,R_p|M_\star)}/\sigma_{f(P,R_p|M_\star)}$ instead of just $\sigma_{f(P,R_p|M_\star)}$ is because while the latter is low for regions where there is a high density of data points, it can also be low where there are no data points (and thus both the mean and standard deviation approach zero). In other words, the ratio $\mu_{f(P,R_p|M_\star)}/\sigma_{f(P,R_p|M_\star)}$ can be thought of as a measure of the significance of the mean probability density relative to its variation arising from the finite sample size. In this example, we note that while the ratio peaks in the region near $P \sim 20$ days and $R_p \sim 2.5 R_\oplus$ (i.e. where there is a high density of planets consisting of the sub-Neptunes above the radius valley), it also peaks in the region of the radius valley \textit{itself} ($P \sim 10$ days and $R_p \sim 1.8 R_\oplus$). This implies that even though there is a reduced occurrence of planets in the radius valley (the mean probability density is low), the radius valley itself is \textit{robust} (the standard deviation of the probability density from the bootstraps is even lower).

The ratio $\mu_{f(P,R_p|M_\star)}/\sigma_{f(P,R_p|M_\star)}$ can also be used to appropriately mask out the regions of where there are too few measurements to provide a robust estimate of the sample density, as shown in the right panel of Figure \ref{fig:CKS-X_period_radius_cond_stmass_bootstraps} (where we show the mean joint probability density where $\mu_{f(P,R_p|M_\star)}/\sigma_{f(P,R_p|M_\star)} > 3$ is chosen for the mask). One has the flexibility to choose the threshold for $\mu_{f(P,R_p|M_\star)}/\sigma_{f(P,R_p|M_\star)}$ depending on how much one wishes to restrict their analyses to regions that are well characterized by the data.
The example here illustrates that a choice of ``$3\sigma$'' for the bootstrap mean can effectively mask out the regions where the data is not well sampled.
Further, one can eliminate the influence of the poorly sampled regions when making predictions with the model (e.g., when computing the mean prediction marginalized over a given dimension) by multiplying the joint distribution with the mask and renormalizing.
A similar procedure for masking out regions of high uncertainty due to finite sample size can be applied for joint probability distributions conditioned on other dimension(s).

\bibliography{references, ManualReferences}

\end{document}